\documentclass{aa}

\usepackage{amsmath}
\usepackage{txfonts}
\usepackage{float}
\usepackage{graphicx}
\usepackage{subcaption}
\usepackage{longtable}
\usepackage{natbib}
\usepackage{multirow}
\usepackage{comment}
\bibpunct{(}{)}{;}{a}{}{,}

\usepackage[colorinlistoftodos]{todonotes}
\usepackage[colorlinks=true, allcolors=blue]{hyperref}
\usepackage{siunitx}

\newcommand{\rms}{r{.}m{.}s{.}~}
\newcommand{\resp}{resp{.}}

\newcommand{\eqs}{Eqs{.}~}
\newcommand{\tab}{Table~}  
\newcommand{\fg}{Fig{.}~}
\newcommand{\fgs}{Figs{.}~}
\newcommand{\sct}{Sect{.}~}
\newcommand{\scts}{Sects{.}~}

\definecolor{darkgreen}{rgb}{0.0,0.5,0.0}
\definecolor{darkred}{rgb}{0.7,0.0,0.0}
\definecolor{brown}{rgb}{0.65,.35,0.}
\definecolor{grey}{rgb}{0.4,0.5,0.55}
\definecolor{royalblue}{rgb}{0,0.5,1}
\definecolor{violet}{rgb}{0.5,0,0.7}
\definecolor{lightgrey}{rgb}{0.85,0.9,0.95}

\usepackage{soul}

\begin{document}

\title{Color dichotomy and gradients in the bulges and disks of EFIGI galaxies along the Hubble sequence}
   \author{Quilley, L.$^{1}$\thanks{email: louis.quilley@univ-lyon1.fr} 
   Lehnert, M. D.$^{1,2}$,
   de Lapparent, V.$^{2}$
   }

\titlerunning{Bulge and disk color dichotomy and gradients along the Hubble sequence}
\authorrunning{Quilley, Lehnert \& de Lapparent (2025)}

\institute{Universit\'e Lyon 1, ENS de Lyon, Centre de Recherche Astrophysique de Lyon (UMR5574), 69230 Saint-Genis-Laval, France \and Institut d'Astrophysique de Paris (UMR7095), Sorbonne Universit\'e, 98 bis boulevard Arago, 75014 Paris, France}
  
   \date{Received 21 August 2025 / Accepted}

\abstract{One of the most outstanding questions in contemporary astrophysics is: What is the significance of galaxy morphology? What physical processes underlie the morphologies we observe and is a galaxy’s internal structure shaping its evolution (nature) or is it a sign of the external processes which drive galaxy evolution (nurture)?}
{We aim to understand the color dichotomy and gradients in bulges and disks along the Hubble sequence.}
{We fit S\'ersic functions to the 2D light distributions in the $g$, $r$, and $i$ bands to bulges and disks of the large EFIGI (Extraction de Formes Id\'ealis\'ees de Galaxies en Imagerie) sample of galaxies with high-quality morphological classifications.}
{In early-type galaxies, bulges and disks have similarly red and nearly uniform colors. Disks become significantly bluer with increasing lateness of their types and bulges get slightly redder because of patchy dust. Disks have increasingly blue colors with increasing radius, whereas dust extinction and scattering leads to smaller effective radii of the bulges and lower steepness of the best-fit S\'ersic functions in $g$ versus $i$. This impact depends on Hubble type, with the bulges of intermediate-type spirals (Sb-Sc) having the reddest mean colors, the largest scatter in their colors, and the largest mean and scatter in their color gradients. Similarly to bulges, disks of the intermediate-type galaxies show the strongest color gradients. The variations in bulge gradients appear to result from the change in the Hubble type rather than in the total galaxy stellar mass.}
{We interpret these properties of the bulges and disks of intermediate type spirals as being due to dust extinction and scattering, which we hypothesize to be an indicator of the gas content and inflow of gas. If early-type galaxies do evolve from massive spiral galaxies, typically intermediate-type spirals, these color gradients are signs of in situ stellar growth within their bulges, likely driven by morphological structure in their disks. These results favor secular evolution (nature) as the primary driver of galaxy evolution in the local Universe.}

\keywords{Galaxies: evolution -- Galaxies: bulges  -- Galaxies : elliptical and lenticular, cD -- Galaxies : spiral -- Galaxies: structure}

\maketitle

\section{Introduction\label{sct-intro}}

2026 marks the 100th anniversary of the Edwin Hubble's discovery that galaxies could be classified by their morphologies into relatively few simple types \citep{Hubble-1926-extragalactic-nebulae}. Over the next 50 years, astronomers realized the significance of a finer classification scheme which added several types \citep{De-Vaucouleurs-1959-class-morph}. This discovery of the regularity in the morphology of galaxies has led to 100 years of debate about whether galaxies are defined by how they are nurtured and their place in large-scale structure, accretion history, and so on, or by their nature and how their internal structure influences their evolutionary path, or even whether morphology is a defining characteristic beyond dividing galaxies into spirals and lenticulars or ellipticals. 

The Hubble sequence can be seen as a sequence of decreasing prominence of the bulge: it is, with the winding of the spiral arms, a criterion with which to differentiate spiral galaxies, whereas ellipticals, as spheroids, can be seen as ``pure bulge,'' and irregulars with no bulge and properties similar to late-type spirals as ``pure disk.'' It has now become a standard approach to analyze the morphology of galaxies by performing a two-component luminosity mode-fitting of their light profiles, known as bulge and disk decomposition \citep{Allen-2006-MGC-BD-decomp, Simard-2011-BD-decomp-SDSS, Lange-2016-bulge-disk-decomposition-GAMA, Margalef-Bentabol-2016-bulge-disk-decomposition-CANDELS-z-3, Dimauro-2018-bulge-disk-decomposition-CANDELS, Fischer-2019-SDSS-IV-MANGA-morpho-catalog, Gao-2019-CIGS-multicomponent-decompositions, Casura-2022-bulge-disk-decomposition-GAMA, Hashemizadeh-2022-DEVILS-bulge-emergence-since-z-1, Genin-2025-DAWN-JWST-morpho-catalog}. This method allows us to obtain separate properties for bulges and disks of galaxies, and hence to study their formation by jointly analyzing the star formation histories of these two components.

Stellar populations of differing luminosity weighted ages emit in different wavelength ranges -- older stellar populations have redder colors than younger populations of stars for a given metallicity and dust extinction. There are many analyses which have found systematic color differences between bulges and disks of galaxies --  some with modest samples \citep{Mollenhoff-2004-BD-spirals-UBVRI} and others with much larger samples of thousands of galaxies \citep[][]{Head-2014-bulge-disk-colors-ETG-Coma-cluster, Vika-2014-multiband-BD-decomposition-MegaMorph, Fernandez-Lorenzo-2014-bulges-pseudobulges-relics, Kennedy-2016-GAMA-color-gradients-vs-BT-color-BD, Barsanti-2021-bulge-disk-colors-clusters, Casura-2022-bulge-disk-decomposition-GAMA}. The state-of-the-art method is deriving the physical properties of the bulges and disks from their photometry through spectral-energy-distribution fitting \citep[see, e.g.,][]{Robotham-2022-ProFuse}, or through integral-field spectroscopy \citep[see, e.g.,][]{Johnston-2017-BUDDI}. Such studies have shown that bulges are older than the disks hosting them \citep{Gonzalez-Deldago-2015-CALIFA-Hubble-sequence-age-metallicity-gradients, Johnston-2022-BUDDI-SFH-bulge-disk-lenticular, Jegatheesan-2024-BUDDI-SFH-bulge-disk-spiral, Bellstedt-2024-CSFH-present-bulges-disks-profuse}.

Beyond the averaged properties of bulges, disks, and galaxies, analyzing the variations in their stellar populations across their spatial extent can provide further constraints on their stellar mass assembly and star formation histories. These spatial variations in stellar ages, metallicities, and dust content manifest themselves as color gradients in photometric surveys. By comparing the structural parameters obtained when performing luminosity weighted model-fitting across various photometric bands, one can detect such gradients, and therefore gain an insight into the way the different populations and gas are distributed within a galaxy.  Many studies have examined the variations in structural parameters in several bands, which provide an interesting insight into galaxy evolution. \citet{Ko-Im-2005-color-gradients-ETG} found that the effective radii of almost 300 low-redshift galaxies were lower in the $K$ band than in the $V$ band. They also interpreted the weaker variation in $R_\mathrm{e}$ with color in denser environments as being a result of enhanced mixing of the stellar populations due to more frequent mergers \citep[see also][for a larger study]{La-Barbera-2010-color-gradients}. The Galaxy And Mass Assembly (GAMA) survey then built on these results by showing the prevalence of color gradients in all galaxy types with variations in $R_\mathrm{e}$ between 15\% and 50\%, with stronger variations in early-type galaxies \citep{Kelvin-2012-GAMA-sersic-fits, Hausler-2013-multiband-sersic-profile-MegaMorph, Vulcani-2014-GAMA-color-gradients, Kennedy-2015-GAMA-color-gradients-vs-z-lum}. Theses analyses also suggest an increase in the S\'ersic $n$ index by a factor as large as 50\% in the redder bands. Interestingly, similar results have been obtained from nonparametric morphological parameter estimates: \cite{Nersesian-2023-non-param-color-gradients} used integral-field spectroscopy to find that galaxy size increases with wavelength.

However, all these studies constrained color gradients of entire galaxies, and studies of color gradients separating bulges and disks remain scarce. \citet{Mollenhoff-2004-BD-spirals-UBVRI} measured increasing and decreasing effective radii across Johnson bands from $U$ to $I$ for the bulge and the disk, respectively. Using the Euclid Early Release Observations (ERO) of the Perseus cluster, \citet{Quilley-2025-ERO} estimated color gradients in whole galaxies and disks from varying effective radii and S\'ersic index at redshifts $\lesssim0.6$. This study showed that single-S\'ersic gradients result from both the bulge-disk color bimodality (the bulge being redder than the disk) and a blue color gradient within the disk. A resolved view of galaxies to separate bulges and disks thus appears pivotal for studying the color gradients of galaxies.

Similar studies from the ground are best performed on nearby well-resolved galaxies. The EFIGI (Extraction de Formes Idealisées de Galaxies en Imagerie) sample, with its detailed visual morphological classification, provides a large statistical sample of well-determined morphological types across the Hubble sequence. Using bulge and disk decomposition, \citet{Quilley-2022-bimodality} identified disk reddening as a key change along the Hubble sequence and across the Green Valley \citep{Wyder-2007-GALEX-CMD-I}, also concomitant with significant bulge growth. More generally, \cite{Quilley-2023-scaling-relations} demonstrated the reliability of the EFIGI effective bulge and disk radii estimates by remeasuring scaling relations for both components.

Here, we examine in more detail the colors and color profiles of both bulges and disks in a weakly elongated subsample of the EFIGI catalog. In \sct\ref{sct-data} we recapitulate the EFIGI and the parameters derived as part of classifying the galaxies within the EFIGI sample. In \sct\ref{sct-methodo} we present information about how the fitting was done and how the final measurements were cleaned of spurious fits. In \sct\ref{sct-bulge-disk-color} we describe our results for the colors of bulges, disks, and galaxies as a function of morphological type. In \scts\ref{sct-disk-gradient} and \ref{sct-bulge-gradient} we present our findings on the disk and bulge gradients respectively. To frame the discussion and because measuring color gradients have a rich literature, in \sct\ref{sct-compare} we compare our results with those in previous studies, and finally in \sct\ref{sct-discussion} we discuss the implications of our results.

\section{Data \label{sct-data}}

In this analysis, we used the visually classified EFIGI morphological catalog of 4458 nearby galaxies \citep{Baillard-2011-EFIGI}. To construct this catalog, a team of astronomers examined $gri$ color composite images extracted from the Sloan Digital Sky Survey (SDSS), assigning a Hubble type to each galaxy and estimating 16 morphological attributes. These attributes included the inner structures (bulge, bar, spiral arms, rings), the overall shape (inclination-elongation, perturbation), texture (dust, HII regions), and surroundings (stellar contamination, neighboring galaxies). 

The attributes used in the present analysis are: \texttt{Inclination-Elongation} (\texttt{Incl-Elong} hereafter), estimating the apparent inclination of disk galaxies or the elongation for disk-less galaxies; and \texttt{Visible Dust}, which evaluates the strength of the diverse features indicating the visual presence of dust in galaxies in the form of clumpy extincted and reddened regions.  The current analysis is actually limited to the subsample of 3106 galaxies with an \texttt{Incl-Elong} attribute of 0, 1, or 2, as for larger values, corresponding to very inclined and edge-on disks, the bulge modeling is less reliable, in part due to its obscuration by the disk and the dust it may contain. This selection maintains all galaxies with elliptical, compact elliptical and dwarf elliptical morphological types within the studied subsample. In addition, we also removed objects that had high values for the contamination parameter (3 and 4). 

The EFIGI catalog was designed to densely sample the variety of types along the Hubble sequence: it mostly includes galaxies with an apparent diameter of $\ge 1$ arcmin, and hence well-resolved multiband SDSS imaging, and most types have several hundred galaxies each. The EFIGI catalog is therefore not representative of the average relative fractions of types, but is a representative sample of each morphological type in the nearby Universe. This sample is well suited for profile-fitting and analysis of the bulges and disks of galaxies along the Hubble sequence. 

\section{Methodology \label{sct-methodo}}

We performed multiband bulge and disk decomposition of the $g$, $r$, and $i$ images of all EFIGI galaxies, using a S\'ersic profile \citep{Sersic-1963-sersic-model} for the bulge and an exponential profile for the disk using \texttt{SourceXtractor++} 
\citep{Bertin-2020-SourceXtractor-plus-plus, Kummel-2020-SourceXtractor-plus-plus}. In this analysis, both model components have elliptical symmetry. The S\'ersic profile has several parameters that we focused on in our analysis: the effective radius, $R_\mathrm{e}$, is the radius that encloses half of the total light; $n$, the S\'ersic index, characterizes the steepness of the profile (the exponential profile corresponds to $n=1$); and the center of the profile fit. The other parameters in the fitting, such as the position angle of the major axis, $\theta$, and the axis ratio, $q$, are not discussed beyond the priors that are set to constrain them. 

We also performed single-S\'ersic profile fitting for all EFIGI galaxies, which were only used here for Sd, Sdm, Sm, and Im galaxies and that generally have insignificant or no bulge. For these late-type galaxies, the S\'ersic index distribution peaks around 1. These single S\'ersic fits were used in the analysis to characterize the disk properties of Sd to Im types. Although Im galaxies are well modeled by elongated approximately exponential profiles, they are likely prolate systems in 3D \citep[e.g.,][]{Swaters09}, and hence they may not generally have a disk morphology. Single S\'ersic modeling could also be considered more appropriate for elliptical galaxies. However, we decomposed the bulges and disks for elliptical and cD galaxies. This is justified physically given the significant rotation seen in kinematic studies of early-type galaxies \citep{Krajnovic-2008-SAURON-evidence-for-disks-in-fast-rotators, Krajnovic-2011-ATLAS-3D-morpho-kinemetric-features-ETG, Emsellem-2011-fast-slow-rotators, Krajnovic-2013-ATLAS-3D-photo-kine-stellar-disks-in-ETG}. 

The luminosity profile fits were performed with the \texttt{SourceXtractor++} \citep{Bertin-2020-SourceXtractor-plus-plus, Kummel-2020-SourceXtractor-plus-plus}, software that allows us to apply priors individually to the bulge structural parameters, as well as using priors on the relative variations of the bulge or disk parameters across the three bands \citep{Quilley-2022-bimodality, Quilley-2023-scaling-relations}. As is explained in {\citet{Quilley-2023-PhD-thesis}, the method starts by fitting the bulge, in a small region centered on the bulge, with a single S\'ersic profile. This fitting process is repeated in the $g$, $r$, and $i$ bands separately. The median of the best fit parameters of the bulge in the three bands is then used as the initial values and centers of the flat prior interval for the corresponding bulge properties in the multiband bulge and disk decomposition (the same ranges apply to the three bands). Comparisons of the fits with and without priors show, in particular, that constraining the fits of the bulges with priors has the significant advantage of preventing one from fitting a hypothetical (and likely unrealistic) elongated bulge component to a possible bar.

Gaussian priors between the $r$ and $i$ bands and the $g$ band fits are used in the decomposition, whose goal is to avoid catastrophic failures, while allowing variations between the different bands to achieve the best fits in each of them. These priors are applied to the logarithm of the ratio for $R_{\rm e}$ and for $q$, to the difference of $x$ and $y$ coordinates of the profile center, of $n$, and of $\theta$, with all 0 means, and standard deviations of 0.1 for $R_{\rm e}$, $q$, and $n$, 0.3 pixels each for $x$ an $y$, and $45\deg$ for $\theta$, respectively. The centering and $n$ priors only apply to the bulge, whereas other priors are used to constrain the variations in the best profile fits to bulge and disk in each galaxy and band. The Gaussian band-to-band priors applied to the single-S\'ersic fits are identical to those applied to the S\'ersic profile for the bulge \citep{Quilley-2023-PhD-thesis}.

A two-stage cleaning is applied to the bulge-disk decompositions, in order to automatically discard spurious fits. In the first stage, we cleaned the $B/T$ values, by fitting in each of the three bands a second degree polynomial to the distributions of the bulge-disk $B/T$ versus the prior $B/T$ (the flux of the prior bulge model divided by the total flux of the galaxy). The distribution within the bulge-disk $B/T$ versus the prior $B/T$ shows an approximately one-to-one sequence, plus data that deviates from this sequence to a much larger bulge-disk $B/T$ than the estimated prior values. These fits were performed using the ODRPACK Version 2.01 Software for Weighted Orthogonal Distance Regression (ODR hereafter, \citealt{ODR-1992}) within the \textit{scipy} \textit{Python} library \citep{2020SciPy-NMeth}, and hence minimize the smallest (instead of vertical in linear regression) distance between the fit and the data points. An iterative $2.5\sigma$ rejection was performed in order to reject the erroneous fits (based on visual inspection of the images of galaxies within various regions in this plane), which converges in eight, six, and seven iterations in the $g$, $r$, and $i$ bands, respectively. A galaxy was rejected if at least one fit in any of the three bands was rejected by this procedure. This is relatively conservative as it requires the fits to be reliable in all bands, as we are interested in both the median values and the dispersion of the S\'ersic parameters as a function of Hubble type. This rejects from $0.5\%$ and $1.5\%$ to $2.7\%$ objects for E and S0$^-$ to S0$^+$ types \resp, $10.3\%$ for S0a types, and $6.5\%$ to $32\%$ from Sa to Scd types (a total of 334 galaxies were rejected). 

In the second stage, we removed galaxies that have fits that are in significant disagreement between the three bands. This was done by comparing the distributions of the logarithm of the maximum to minimum values in the three bands for the bulge and disk effective radii, of the difference between the maximum and minimum bulge and disk axis ratios, and the normalized distance (with the component effective radii) between the bulge and disk model centers. We rejected the fits for the 59 galaxies of Scd types and earlier that have values beyond three times the 90\% percentiles of at least one of these six distributions. For types Sd and later that are modeled by a single-S\'ersic profile, the same process was applied to the effective radii and axis ratio distributions, and led to the removal of 14 single-S\'ersic fits. The cleaned sample contains 1824 bulge-disk decomposition for types earlier than Scd, and 684 single-Sérsic models for types Sd to Im.

All apparent magnitudes measured by \texttt{SourceXtractor++} were corrected for  atmospheric extinction, using the $kk$ coefficients multiplied by the air masses provided for the SDSS\footnote{\url{https://classic.sdss.org/dr5/algorithms/fluxcal.html}}. The correction for galactic extinction was based on the maps of galactic dust reddening, from which  $E(B-V)$ values were obtained for each galaxy using its sky coordinates, which we converted to extinction in the $g$, $r$, and $i$ bands \citep{SFD-1998-gal-extinction}. We used the conversions to extinction factors \citep{S-and-F-2011-gal-extinction} assuming an extinction to reddening ratio of $A_V/E(B-V)=3.1$.

Finally, these extinction-corrected apparent magnitudes and HyperLeda redshifts were used to fit spectral energy distributions (SEDs) with the ZPEG software \citep{Le-Borgne-2002-ZPEG}. From these fits, we derived the absolute (rest-frame) magnitudes and colors, as well as stellar masses}. Only dust-free templates were used as $g$, $r$, and $i$ photometry alone does not allow one to properly constrain dust extinction. This SED-fitting process was performed first for the magnitudes of whole galaxies \citep[complemented with $NUV$ fluxes measured with GALEX for 1040 entire galaxies among the 2508 galaxies in the present sample;][]{Bianchi-2017-GALEX-GUVcat-AIS}. It was then repeated separately for the magnitudes of the bulges and disks, in order to obtain the absolute magnitudes, and hence colors, of these components in the $g$, $r$, and $i$ bands \citep[see][for details]{Quilley-2022-bimodality}. Most notably, bulges and disks were shown to be best fit by templates matching their specific colors as a function of the Hubble type.

\begin{figure*}[ht]
\centering
\includegraphics[width=\linewidth]{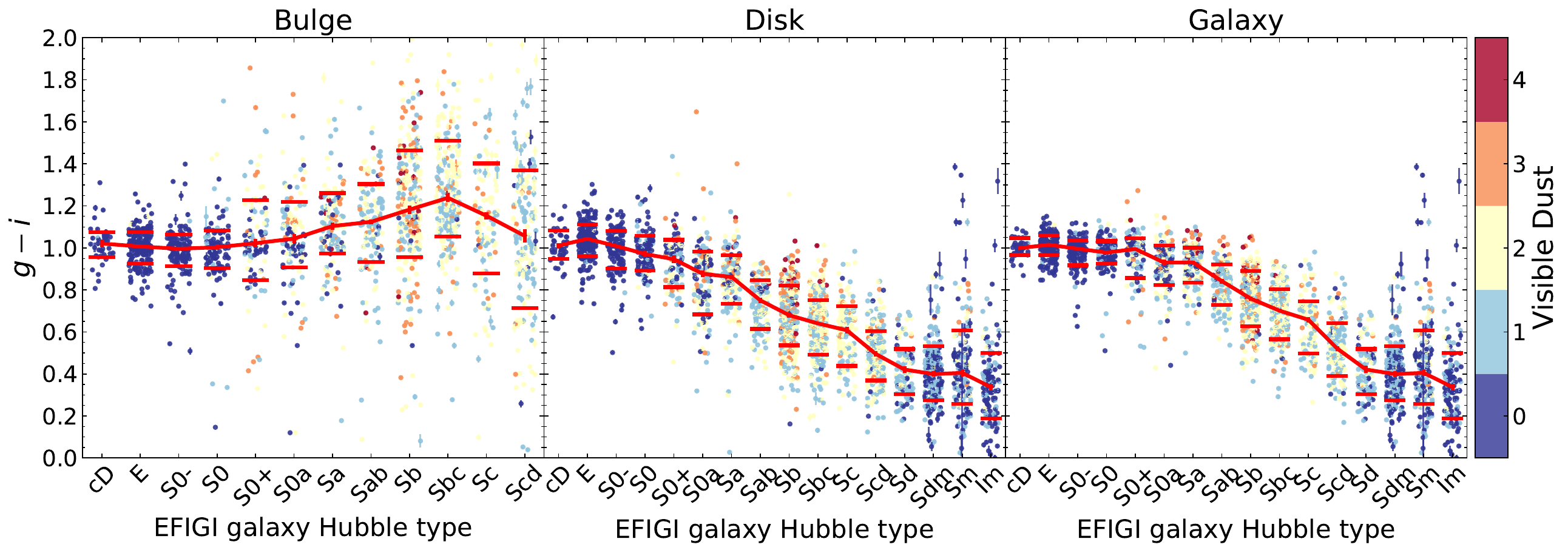}
\caption{Distributions of the $g-i$ absolute colors of the bulges (left), disks (center), and full galaxies (right) from the bulge-disk decompositions as a function of morphological type of EFIGI galaxies with {\tt Incl-Elong}$\leq 2$. The points are color-coded by the EFIGI attribute \texttt{Visible Dust} and the uncertainties in each points were derived by adding in quadrature the uncertainties estimated by \texttt{SourceXtractor++} for both apparent magnitudes (\sct\ref{sct-methodo}). The results for galaxies and disks with Sd-type and later are shown for the single-S\'ersic modeling only. The red lines indicate the median $g-i$ color per type, with the vertical error bars representing the associated bootstrap uncertainties in the median values, and the horizontal segments above and below correspond to the 16th and 84th percentiles of the color distributions (see \tab \ref{tab-colors-gi}). Disks show a systematic bluing with increasing Hubble type, whereas spiral bulges become on average redder and their colors more dispersed from early- to intermediate-type spirals.}
\label{fig-color-B-D}
\end{figure*}

Aperture photometry has been used frequently as the direct and simplest way of measuring colors of galaxies: it encloses a fixed region of each object, provided that the point spread functions are similar in all bands. The circular apertures used in this approach are best suited for round and thus generally face-on spirals, and tend to generate biases when observing a given galaxy at different disk inclinations. Here, we chose to estimate the colors of galaxies, as well as their bulges and disks in the EFIGI sample, by using the model magnitudes of the fit profiles (integrating the fit light profile to infinity). The colors resulting from model magnitudes are identical to those for the bulge and disk truncated at their respective effective radii, as the corresponding fluxes are half the model flux, and hence within different elliptical apertures, compared to a common circular one for colors from aperture photometry. The models are mostly constrained by the highest-surface-brightness parts of each galaxy (i.e., at small radii) and the models are relatively insensitive to low-surface-brightness emission (i.e., at large radii).  Moreover, the data at large radii are noisier and \texttt{SourceXtractor++}  weighs pixels according to their signal-to-noise. \citet{Bernardi-2003-elliptical-colors-color-gradients-chemical-abundances} showed that for elliptical galaxies the model colors are those with the weakest dependence on the absolute magnitude. An intermediate approach between the model and aperture colors would be to integrate the models within a common elliptical aperture in the different bands. Another approach would be to use a common model in all bands, and hence identical $R_{\rm e}$, $n$, and elliptical apertures. We prefer, however, to allow for some differences between bands, in order to detect and model subtle effects such as color gradients. 

\section{Bulges and disks colors across Hubble types\label{sct-bulge-disk-color}}

Figure \ref{fig-color-B-D} shows the $g-i$ absolute colors of bulges (left), disks (center), and full galaxies (right) as a function of their Hubble type. All values are listed in \tab \ref{tab-colors-gi}. The $g$ and $i$ bands are the bluest and reddest SDSS bands available that have sufficient signal-to-noise with which to conduct our analysis.  Bulges are only shown for Scd and earlier types. The colors of bulges (left panel of \fg\ref{fig-color-B-D}) show an increase in the median color of $0.12$ magnitudes from $g-i=1.00$ for ellipticals to the peak value of $1.12$ for Sbc intermediate-type spirals. Another striking feature of the bulge color distribution with type is the marked increase in the \rms dispersion of the bulge colors per type by a factor of $\approx2-3$ for S0$^+$ to Scd types (in the $0.23$ to $0.34$ interval) compared to that for E and S0$^-$ ($0.08$ and $0.11$ \resp , see column ``SD'' of \tab\ref{tab-colors-gi}).
Estimating the uncertainties with a bootstrap method in the median colors, the median bulge colors are $0.7\sigma$, $2.9\sigma$, and $5.1\sigma$ redder for Sab, Sb, and Sbc types, respectively, than for Sa galaxies. Bulges of Sc and Scd types are on average bluer than Sbc types by $0.08$ and $0.18$ mag, respectively. These color decreases are significant at the $3.2\sigma$ and $4.7\sigma$ levels, respectively.

By contrast, the central panel of \fg\ref{fig-color-B-D} shows a progressive $0.62$ mag bluing in the median disk color from $1.0$ mag for E types to $\approx0.4$ mag for Sd, Sdm, and Sm types and an insignificant increase in the dispersion. This bluing is statistically significant -- at the $8\sigma$ level between the adjacent Sa and Sab types. Overall, the mean disk color decreases by $13\sigma$ from S0$^-$ to Sa, by $17\sigma$ from Sa to Sbc types, and by $12\sigma$ from Sbc to Sd types. These results imply that disk color is an indicator of the Hubble type from early-type galaxies, E, up to disk-dominated Sd galaxies \citep[as already highlighted in][]{Quilley-2022-bimodality}. At later types after Sd galaxies, mean colors become roughly constant but with a small increase in the \rms dispersion (\tab\ref{tab-colors-gi}).

Finally, the right panel of \fg\ref{fig-color-B-D} shows that the total galaxy colors follow a similar bluing with Hubble type as that measured for the disks, from $1.01$ mag (for E types) to $0.42$ mag (for Sd types), then a flattening with similarly blue colors for Sdm and Sm types. The decrease in color in galaxies along the Hubble sequence (from early to late types) is the result of both the disks becoming increasingly blue, as well as the decreasing flux contribution of the (generally redder) bulges (see \citealt{Quilley-2022-bimodality}). The combination of similar colors of the bulge and disk components in early-type galaxies leads to negligible change in the mean colors of galaxies between E and S0 types. Similarly, both components of cD types, and hence whole galaxies, have similar colors as the E and S0 types.
Overall, similar trends are observed for the bulge, disks, and whole galaxies in the narrower $g-r$ color range (\tab \ref{tab-colors-gr}).

\begin{figure*}[ht]
\centering
\includegraphics[width=\linewidth]{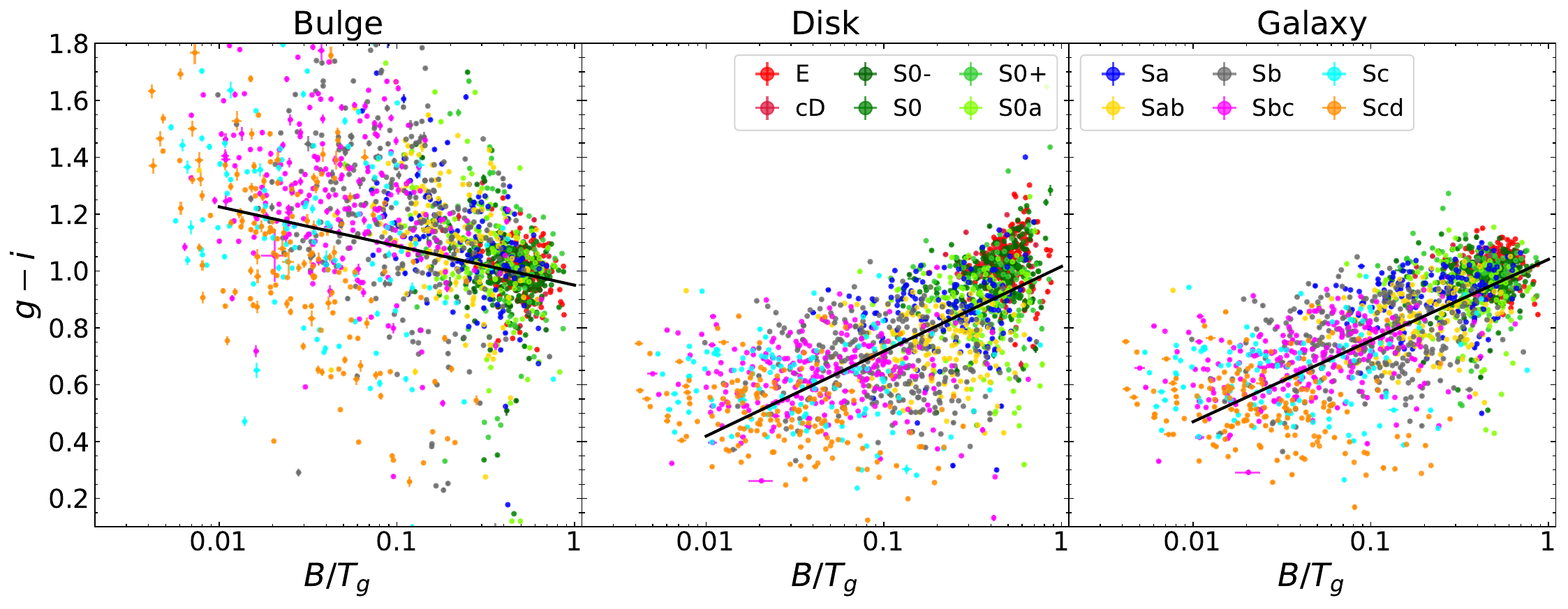}
\caption{$g-i$ colors of the bulge (left), disk (center), and full galaxies (right) as a function of $B/T$ estimated in the $g$ band, from the bulge-disk decompositions of all EFIGI galaxies with \texttt{Incl-Elong}$\leq 2$. The points are color-coded with the EFIGI Hubble type. The solid black lines are linear fits of $g-i$ as a function of $\log({B/T}_g)$ to galaxies with ${B/T}_g > 0.01$ (\eqs\ref{eq-gi-bulge-color-BT} to \ref{eq-gi-galaxy-color-BT}).}
\label{fig-color-B-D-vs-BT}
\end{figure*}

\citet{Quilley-2022-bimodality,Quilley-2023-scaling-relations} showed that the bulge-to-total ratio, $B/T$, increases monotonously along the Hubble sequence from late to early types, and that the position of galaxies across some key evolutionary plots, namely the color-magnitude diagram, the size-luminosity relation, and the Kormendy relation, is predominately related to $B/T$. Therefore, we also examined the variations in the bulge, disk, and full galaxy colors as a function of $B/T$ (\fg\ref{fig-color-B-D-vs-BT}). This confers the advantage of allowing for future comparisons to be made with galaxy surveys at various redshifts, which likely will not have a visually classified Hubble type as in EFIGI, but whose morphology can be parametrized with $B/T$. 

In the left panel of \fg\ref{fig-color-B-D-vs-BT}, the bulge color shows a mild variation with ${B/T}_g$, with the lower ${B/T}_g < 0.1$ of intermediate spiral types having slightly redder colors than the more prominent bulges (${B/T}_g \sim 0.4$) of lenticulars. In contrast, both disk (central panel) and full galaxies (right panel) become increasingly red as the bulge becomes more prominent and exhibit steeper color variations with ${B/T}_g$. These variations are quantified by a linear fit of all three $g-i$ colors as a function of $\log({B/T}_g)$ for galaxies with ${B/T}_g > 0.01$.
We obtain the following relations:
\begin{equation}
    (g-i)_\mathrm{bulge} = (-0.138\pm0.013) \log({B/T}_g) + (0.950\pm0.013)
    \label{eq-gi-bulge-color-BT}
,\end{equation}
\begin{equation}
    (g-i)_\mathrm{disk} = (0.298\pm0.008) \log({B/T}_g) + (1.016\pm0.007)
    \label{eq-gi-disk-color-BT}
,\end{equation}
\begin{equation}
    (g-i)_\mathrm{galaxy} = (0.285 \pm 0.006) \log({B/T}_g) + (1.041 \pm 0.006)
    \label{eq-gi-galaxy-color-BT}
.\end{equation}
The Pearson correlation coefficients are 0.71 and 0.67 for full galaxies and disks, respectively, indicating a noticeable correlation, but is -0.23 for bulges, indicating that the observed anticorrelation is much less significant. This is further confirmed by the $R^2$ scores of the fits, which quantify the proportion of the variation of $g-i$ that is predictable from $B/T$, and are 0.50 and 0.45 for galaxies and disks, respectively, but only 0.05 for bulges. Finally the \rms dispersions around the fits are 0.14 and 0.17 for galaxies and disks, respectively, against a higher value of 0.29 for bulges. Equations~\ref{eq-gi-bulge-color-BT} to \ref{eq-gi-galaxy-color-BT} can be rewritten as a function of $B/T$ in the $r$ or $i$ bands using \eqs\ref{eq-BTi-vs-BTg} and \ref{eq-BTr-vs-BTg} (see also \fg\ref{fig-BT-var-gi}).

\begin{figure}[ht]
\centering
\includegraphics[width=\columnwidth]{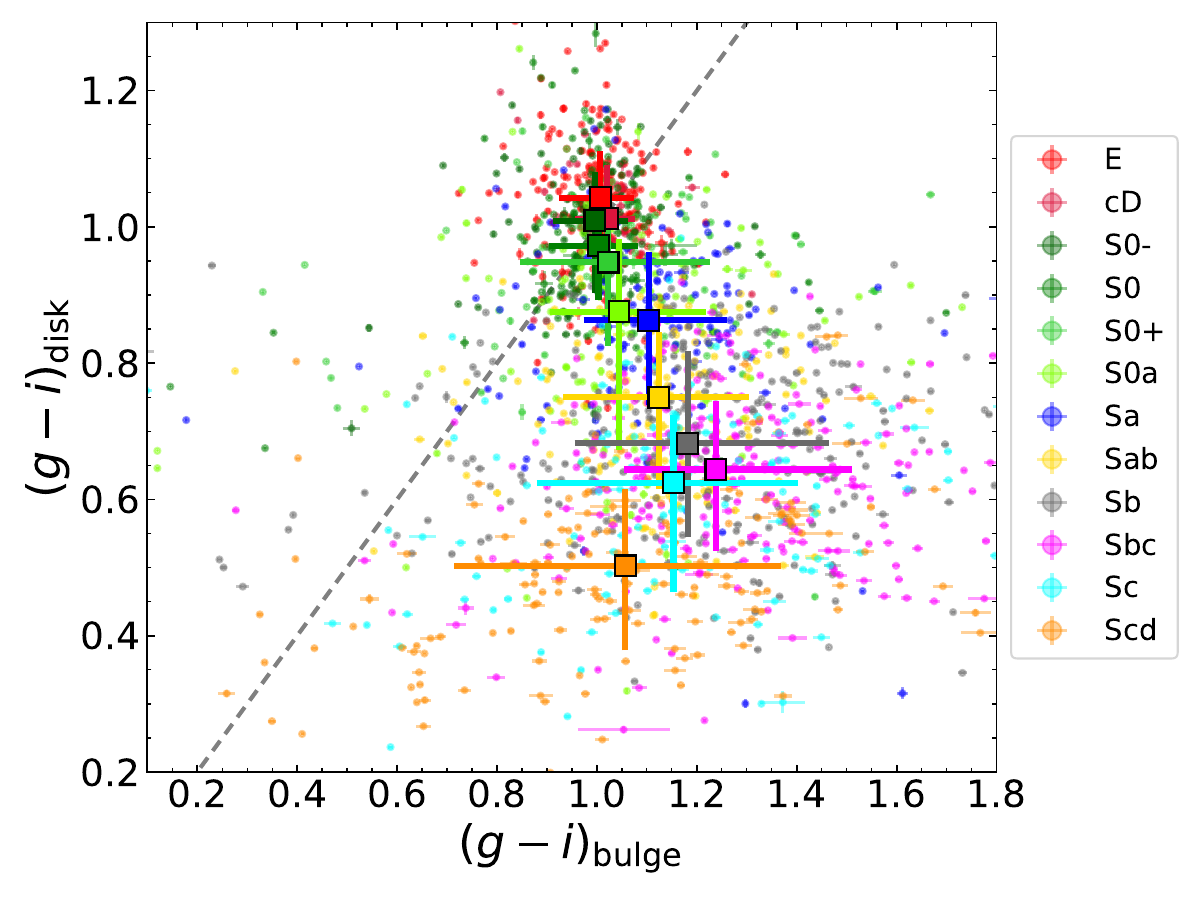}
\caption{Disk versus bulge $g-i$ color for our sample; color-coded by Hubble type (as indicated in the legend on the right). The median value for each Hubble type is indicated by the colored squares, and the 16th to 84th percentiles of the color distribution by the horizontal and vertical bars.}
\label{fig-color-D-vs-B}
\end{figure}

In \fg\ref{fig-color-D-vs-B}, we directly compare the $g-i$ bulge and disk colors of all galaxies types along the Hubble sequence in which the bulge and disk components are reliable (hence types Scd and earlier). Almost all spiral galaxies are located right of the dashed black line representing an identical $g-i$ bulge and disk color, as they have redder bulges than disks. Progressively, earlier-type galaxies have increasingly similar bulge and disk colors. Specifically, E and S0 types have similar bulge and disk colors, with a larger color dispersion for bulges of types S0+ (\fg\ref{fig-color-B-D} and \tab\ref{tab-colors-gi}). The distributions of the difference between the bulge and the disk $g-i$ colors of EFIGI galaxies as a function of both Hubble type and ${B/T}_g$ are plotted in \fg\ref{fig-color-diff-BD} and listed in \tab \ref{tab-colors-gi}.

\section{Disk color gradients along the Hubble sequence\label{sct-disk-gradient}}

\begin{figure}[ht]
\centering
\includegraphics[width=\columnwidth]{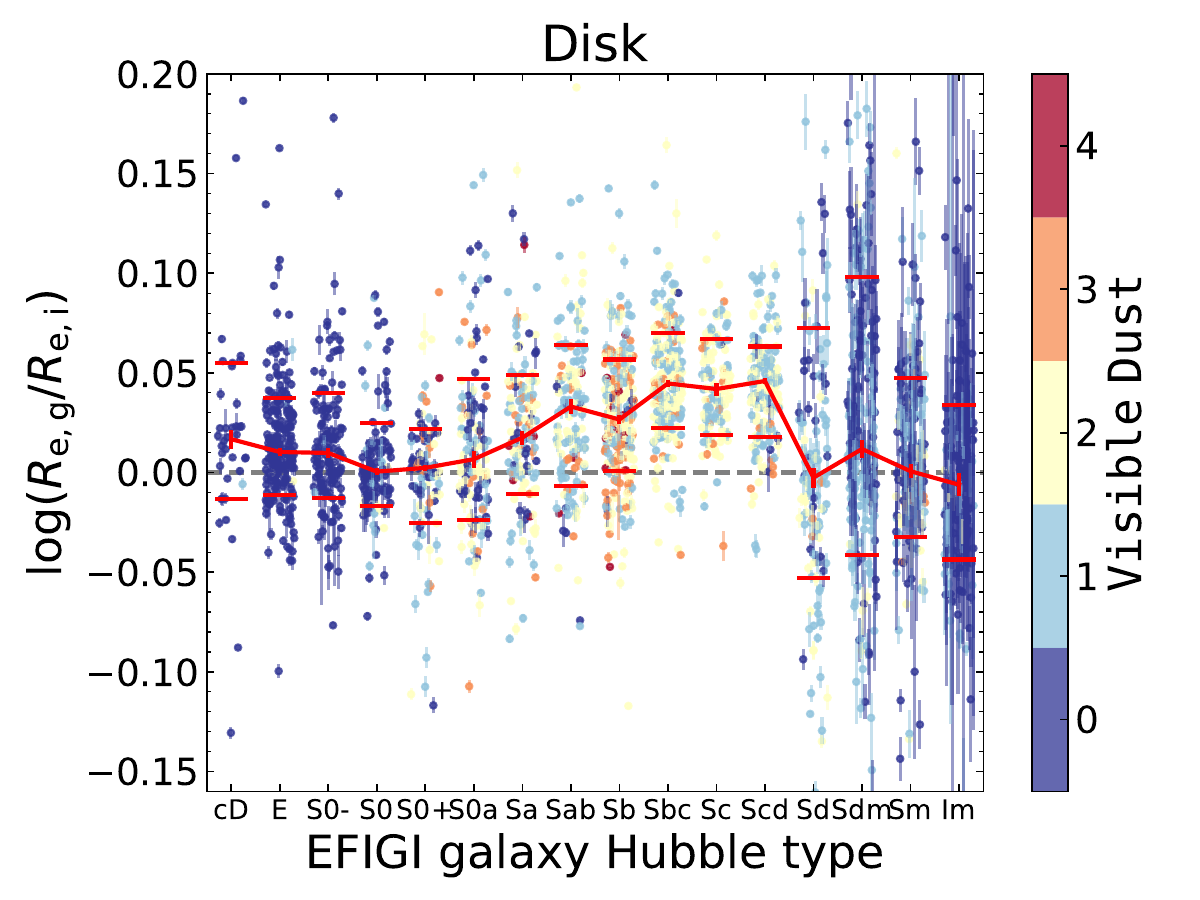}
\caption{Relative distribution of effective radii, $\log(R_{\mathrm{e},g}/R_{\mathrm{e},i})$, of the disk components of the bulge-disk decompositions as a function of Hubble type. The points are color-coded according to the dust content of the galaxy, estimated by the visual attribute \texttt{Visible Dust}. The solid red line shows the median values per type (with associated bootstrap uncertainties) and the horizontal segments the 16th and 84th percentiles of the distributions per type. Early- and intermediate-type spirals exhibit significant color gradients in their disk. These gradients mean that disks are generally bluer with increasing radius.}
\label{fig-color-gradients-disks}
\end{figure}

Figure \ref{fig-color-gradients-disks} shows the logarithm of the ratio between the effective radii of the EFIGI disk components in the $g$ and $i$ bands, $R_{\mathrm{e},g}$ and $R_{\mathrm{e},i}$ respectively, as a function of Hubble type. All values are listed in \tab \ref{tab-gradients-gi}. We color-coded each galaxy according to its ``visual dust content'' (the \texttt{Visible Dust} attribute in the EFIGI catalog). This attribute measures the strength and frequency of indications of the presence of dust, from diffuse dust causing overall reddening, to dust clouds and lanes that can be identified by lowering the surface brightness and reddening the color of particular regions. Except for Sd and later-type galaxies, all median $R_{\mathrm{e},g}/R_{\mathrm{e},i}$ ratios are above 1, suggesting blue radial color gradients in the disks (but not uniquely as there are a minority disks of all galaxy types which show red color gradients). An examination of galaxy data, model and residual images indicates that parts of these color gradients are contributed to by structures within the disks with specific colors and within specific radii ranges, such as bars, spiral arms, rings, and lenses.

The dispersion of the disk color gradients per type (\fg\ref{fig-color-gradients-disks}) lies in the narrow $0.02$--$0.05$ dex interval for all galaxy types down to Scd (see \tab\ref{tab-gradients-gi}), allowing us to detect small differences in effective radii between the $g$ and $i$ bands. For E and S0$^-$ types, the median effective radii of the fit to the disk in the $g$ band is larger than for the fit to the $i$ band by $2.3\%$ (this is further discussed in \sct\ref{sct-bulge-gradient-dispersion}). For S0 and S0$^+$ lenticular galaxies, the mean ratios between the two effective radii indicate no significant difference between the two bands. For spiral types later than S0a, the median $R_{\mathrm{e},g}$ is larger than $R_{\mathrm{e},i}$ by $4$\%, $6$\%, and $10$\% for Sa, Sb, and Sc type galaxies, respectively. The median effective radius ratios are therefore significantly larger than those for all types from S0a to Sm, with a significance of $4\sigma$ (for S0a types), 8 (Sa), $26\sigma$ (Sab), $32\sigma$ (Sb), $54\sigma$ (Sbc), $34\sigma$ (Sc), and $55\sigma$ (Scd). We show in \fg\ref{fig-color-gradients-BT} that there is also a variation in the gradients as a function of ${B/T}_g$.

We also measure a local dip in the mean $R_{\mathrm{e},g}/R_{\mathrm{e},i}$ for types Sb when compared to types Sab and Sbc, which interestingly is the galaxy type for which the \texttt{Visible Dust} attribute takes overall and, on average, larger values than for any other Hubble type \citep{de-Lapparent-2011-EFIGI-stats}. In contrast, for Sd to Im types, for which single S\'ersic fits are used (see \sct\ref{sct-methodo}), the median $R_{\mathrm{e},g}/R_{\mathrm{e},i}$ ratios are consistent with no gradient. The larger individual uncertainties for these late-type galaxies, due to their lower surface brightness, increase the scatter and may prevent us from measuring weak color gradients as well as any trend in color gradients.

\section{Complex light distribution of spiral bulges\label{sct-bulge-gradient}}

\subsection{Large dispersion with color in bulge radii and indexes\label{sct-bulge-gradient-dispersion}}

\begin{figure*}[ht]
\centering
\includegraphics[width=0.8\textwidth]{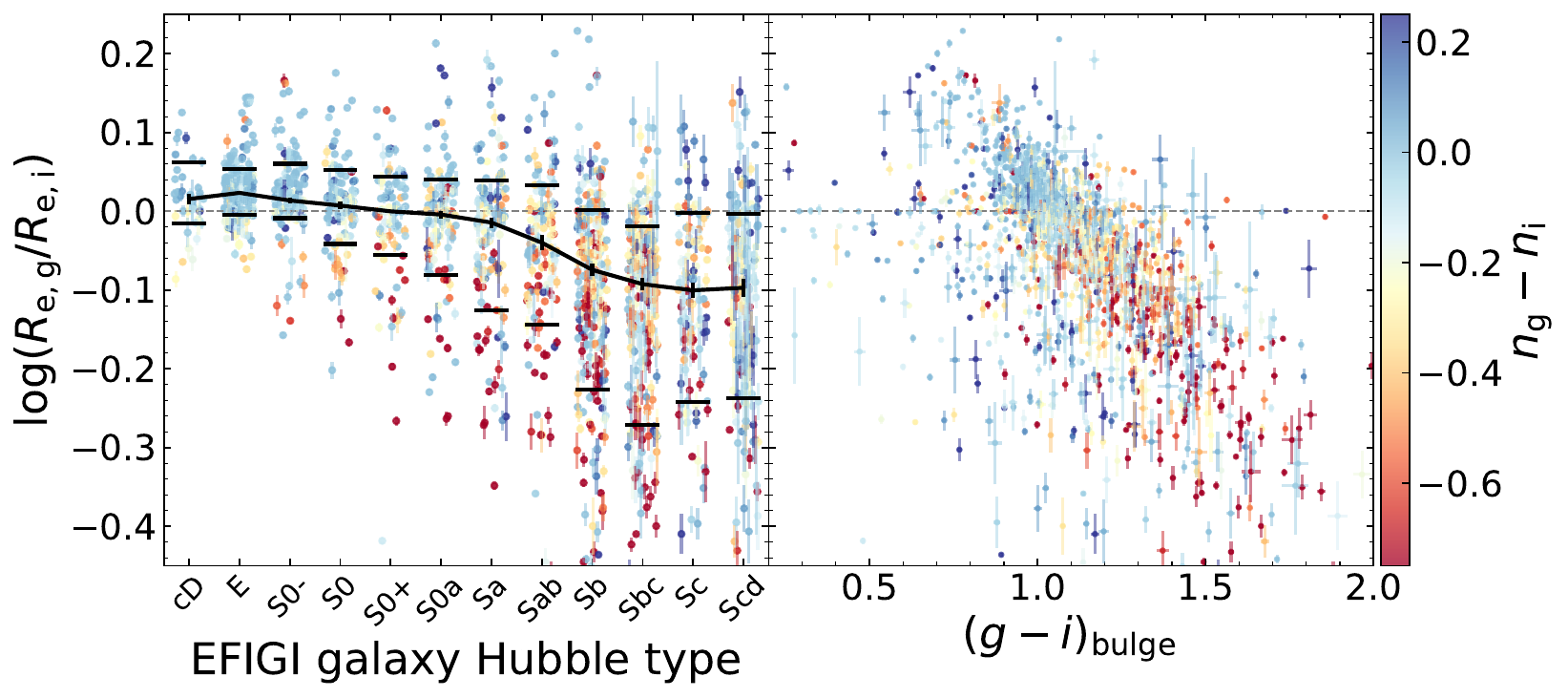}
\caption{$g$-to-$i$ ratio of effective radii (in log), for the bulge components of the bulge-disk decompositions of all types up to Scd, as a function of Hubble type (left) and bulge $g-i$ color (right). Each galaxy is color-coded by its bulge S\'ersic index difference between the $g$ and $i$ bands. The solid line shows the median value per type, the vertical error bars show the associated bootstrap uncertainty in the median, and the horizontal segments show the 16th and 84th percentiles of the distribution. The bulges with the strongest red gradients and the reddest colors tend to have the largest differences in their $g$ and $i$ S\'ersic indices. Galaxies of S0$^-$ type and earlier have predominately weak blue gradients, $g-i$ colors of $\sim1$, and relatively small differences in their bulge S\'ersic indexes.}
\label{fig-color-gradients-bulges}
\end{figure*}

The left panel of \fg\ref{fig-color-gradients-bulges} shows the ratio between $R_{\mathrm{e},g}$ and $R_{\mathrm{e},i}$, the bulge effective radii in the $g$ and $i$ bands, respectively. The bulge components of E, S0$^-$, and S0 types have $R_{\mathrm{e},g}$ larger than $R_{\mathrm{e},i}$ by $5$ to $2$\%, with a small dispersion of $0.04$ to $0.06$ dex. Together with the less than $2\%$ corresponding ratio of the exponential components for these types (see \sct\ref{sct-disk-gradient}), these gradients contrast with the $\approx15\%$ larger effective radii in $g$ than in $i$ measured for ellipticals by \citet{Bernardi-2003-elliptical-colors-color-gradients-chemical-abundances}. The difference may be caused by the fact that these authors use the SDSS fits of a single de Vaucouleurs profile (hence with a S\'ersic index fixed to 4). Moreover these fits were performed independently in each band. In contrast, we find that 90\% of the indexes of the bulge components of our fits to E galaxies are $\le4$ with a peak at $2.9$, in both the $g$ and $i$ bands. The $R_{\mathrm{e},g}/R_{\mathrm{e},i}$ ratio for E galaxies using our single S\'ersic fits has a median value of $0.97\pm0.01$, and hence a negative gradient (only $2.5\sigma$ away from a null gradient), and S\'ersic indexes peaking at $5$--$5.5$ in the $g$ and $i$ bands. The positive gradients measured for both the S\'ersic and exponential components to E types confirm that their fits with a single S\'ersic profile do not capture the complexity of these galaxies (see \sct\ref{sct-methodo}). The variations between bands in the S\'ersic index and effective radii are not independent, and color gradients manifest themselves in both parameters. Therefore, the choice of a fixed S\'ersic index of $n=4$ for all bands by \citet{Bernardi-2003-elliptical-colors-color-gradients-chemical-abundances} forces all the band-to-band variations to be carried by the effective radii.

In \fg\ref{fig-color-gradients-bulges} (left panel), the mean $R_{\mathrm{e},g}/R_{\mathrm{e},i}=1.0$ and $0.99$ for S0$^+$ and S0a types, respectively, with an increasing dispersion of $0.08$ and $0.10$ dex compared to E types. Then for Sa to Scd types the bulges present the opposite trend to the disks (described in \sct\ref{sct-disk-gradient}): the $g$-band effective radii are significantly smaller than those in the $i$ band as Hubble types shift through spirals (by $3$\%, $9$\%, $16$\%, $19$\%, $20$\%, $20$\%, for Sa, Sab, Sb, Sbc, Sc, Scd types, respectively). We verified in \fg\ref{fig-D-vs-B-gradients} that this correspondence between the trends for the bulge and disk gradients is not caused by the modeling. Another characteristic of the bulge effective radius ratio distribution is that the dispersion per Hubble type increases steadily from $\approx0.05$ dex for S0$^-$ galaxies to $\approx0.15$ dex for intermediate spiral types. 

The interpretation of different effective radii for bulges between bands is not straightforward, as one must also take into account the variations in the S\'ersic indexes between bands. Examining the relative distribution of effective radii as a function of Hubble type with a color-coding by the index difference $n_g-n_i$ in the left panel of \fg\ref{fig-color-gradients-bulges} shows that for types S0a through Sc, galaxies with low values of $R_{\mathrm{e},g}/R_{\mathrm{e},i}$ tend to have negative values of the index difference, and hence lower S\'ersic indexes in the $g$ band compared to that in the $i$ band. Also, the lowest values and largest differences in the absolute value of $n_g-n_i$ appear predominantly for galaxies with an effective radius ratio below the mean per type (with the exception of Scd galaxies, the latest spiral type for which bulges are modeled -- but their small values of $B/T$ makes the index determination more uncertain). The smaller indexes of the pseudo-bulges of late spiral types would also make the absolute index differences smaller. The distribution of the index difference $n_g-n_i$ as a function of Hubble type shows the larger index differences and dispersion for Sab, Sb, and Sbc galaxies, compared to earlier and later types (see \fg\ref{fig-sersic-color-gradient}). 

Despite the wide dispersion in the $g-i$ color and $R_{\mathrm{e},g}/R_{\mathrm{e},i}$ ratio for the bulges of spiral galaxies, seen in \fg\ref{fig-color-B-D} and \ref{fig-color-gradients-bulges}, respectively, we also observe in the right panel of \fg\ref{fig-color-gradients-bulges} a striking relationship between the $g-i$ color and the $R_{\mathrm{e},g}/R_{\mathrm{e},i}$ ratio, across all Hubble types: redder bulges have higher bulge effective radii in $i$ than in $g$ (red color gradient). This trend has a large dispersion in $n_g-n_i$ when the bulges are red ($g$-$i$$\gtrsim$1.0) and have a negative radius ratio. In contrast, the positive effective radius ratios have a much smaller dispersion in $n_g-n_i$ and correspond essentially to E and S0 types. For these early types, the $g$ effective radii of the bulge components are therefore on average larger than in $i$, and show little change in the steepness of their profiles with color. We also checked in Appendix~\ref{sct-appendix-bar} that the bulge gradients are not affected by a modeling bias from the presence of a bar.

\begin{figure*}
\includegraphics[width=0.122\textwidth]{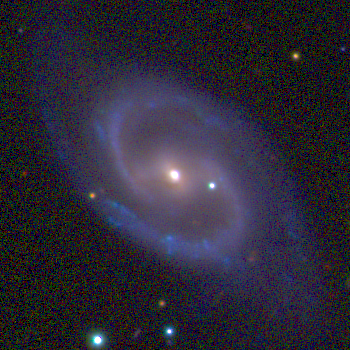}
\includegraphics[width=0.122\textwidth]{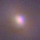}
\includegraphics[width=0.122\textwidth]{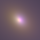}
\includegraphics[width=0.122\textwidth]{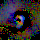}
\includegraphics[width=0.122\textwidth]{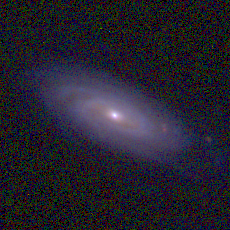}
\includegraphics[width=0.122\textwidth]{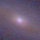}
\includegraphics[width=0.122\textwidth]{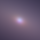}
\includegraphics[width=0.122\textwidth]{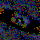}
\includegraphics[width=0.122\textwidth]{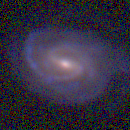}
\includegraphics[width=0.122\textwidth]{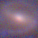}
\includegraphics[width=0.122\textwidth]{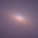}
\includegraphics[width=0.122\textwidth]{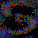}
\includegraphics[width=0.122\textwidth]{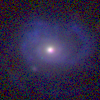}
\includegraphics[width=0.122\textwidth]{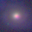}
\includegraphics[width=0.122\textwidth]{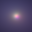}
\includegraphics[width=0.122\textwidth]{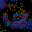}
\includegraphics[width=0.122\textwidth]{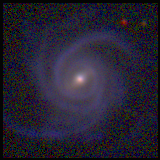}
\includegraphics[width=0.122\textwidth]{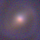}
\includegraphics[width=0.122\textwidth]{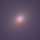}
\includegraphics[width=0.122\textwidth]{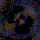}
\includegraphics[width=0.122\textwidth]{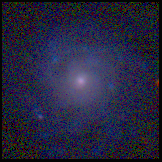}
\includegraphics[width=0.122\textwidth]{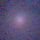}
\includegraphics[width=0.122\textwidth]{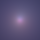}
\includegraphics[width=0.122\textwidth]{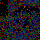}
\includegraphics[width=0.122\textwidth]{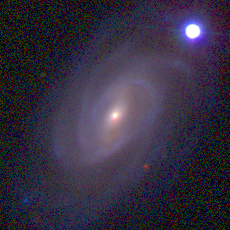}
\includegraphics[width=0.122\textwidth]{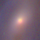}
\includegraphics[width=0.122\textwidth]{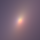}
\includegraphics[width=0.122\textwidth]{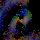}
\includegraphics[width=0.122\textwidth]{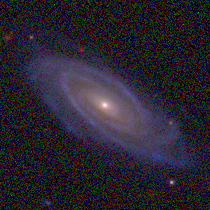}
\includegraphics[width=0.122\textwidth]{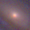}
\includegraphics[width=0.122\textwidth]{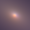}
\includegraphics[width=0.122\textwidth]{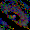}
\caption{From left to right and top to bottom: Images of PGC0023028 (Sb), PGC0027859 (Sab), PGC0029460 (Sb), PGC0033792 (Sb), PGC0033914 (Sc), PGC0039280 (Scd), PGC0047612 (Sbc), and PGC0054232 (Sc), displaying bulge lateral offsets. For each object, one can see from left to right the full galaxy data image, the zoomed-in image of each bulge, the zoomed-in image of the model, and the residual zoomed-in image obtained by subtracting the two previous images.}
\label{fig-bulge-exple}
\end{figure*}

\subsection{Transverse color gradients in bulge images\label{sct-bulge-gradient-shift}}

Understanding the nature of the bulge gradients seen in \fg\ref{fig-color-gradients-bulges} requires a visual examination. The images used for the EFIGI visual classifications cannot be used to examine bulge colors, because their RGB combinations of $i$, $r$, and $g$ images were set to levels that made the low-surface-brightness features of the disks visible. As a result, most bulges appear saturated, masking any visual clues about their colors.

In order to visually examine the colors of bulges, we rescaled the colors levels of a subsample of galaxy images. Specifically, we selected the 69 bulges of galaxies with $-0.2\le\log{R_{\mathrm{e},g}/R_{\mathrm{e},i}}\le-0.1$ and $0.05<B/T(g)<0.1$, which are dominated by Sb to Scd types, as well as 16 Sab types within the same radius ratio interval as the other 69 galaxies, but with $0.1<B/T(g)<0.2$. Figure \ref{fig-bulge-exple} shows a selection of bulge images, in which we observe transverse color gradients across the bulges, and these are well reproduced in the model images. 

\begin{figure*}[ht]
\centering
\includegraphics[width=0.95\columnwidth]{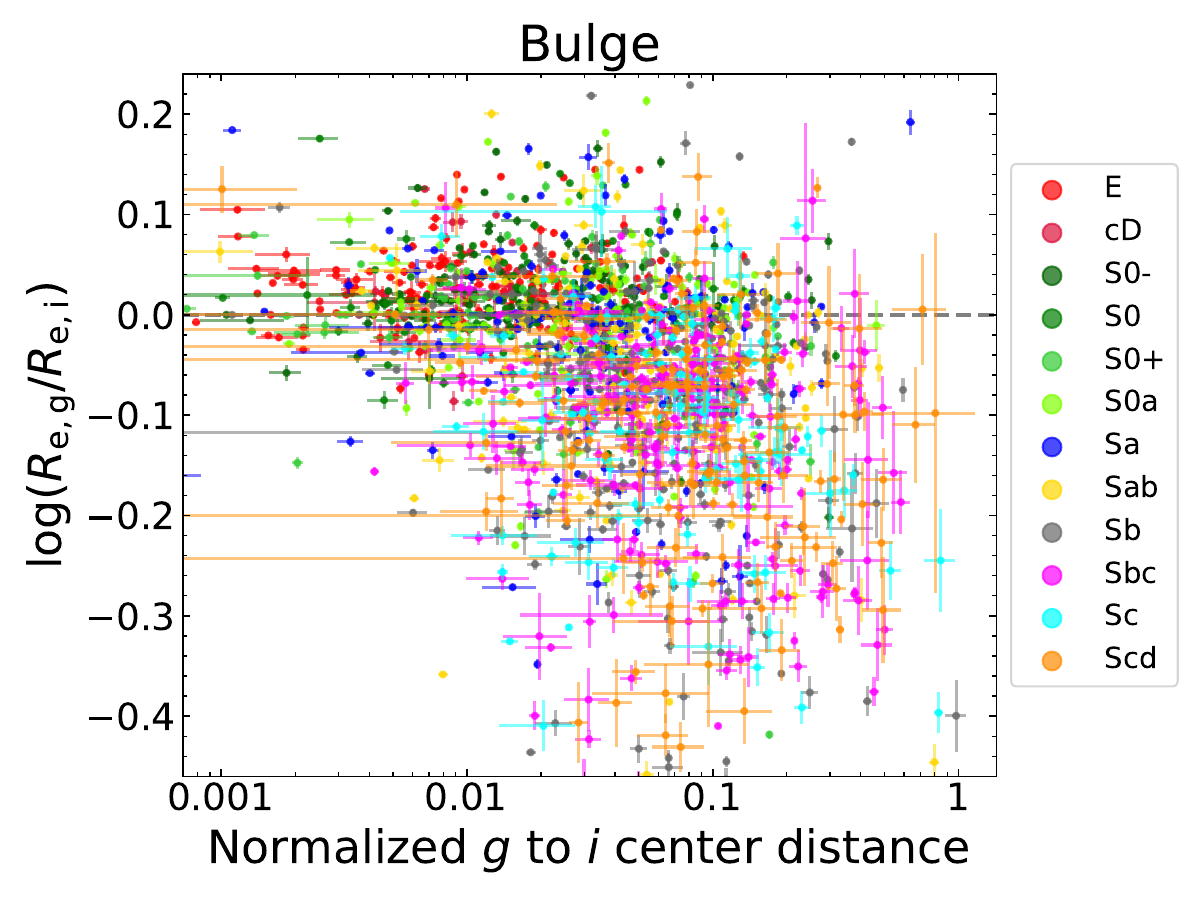}
\includegraphics[width=0.9\columnwidth]{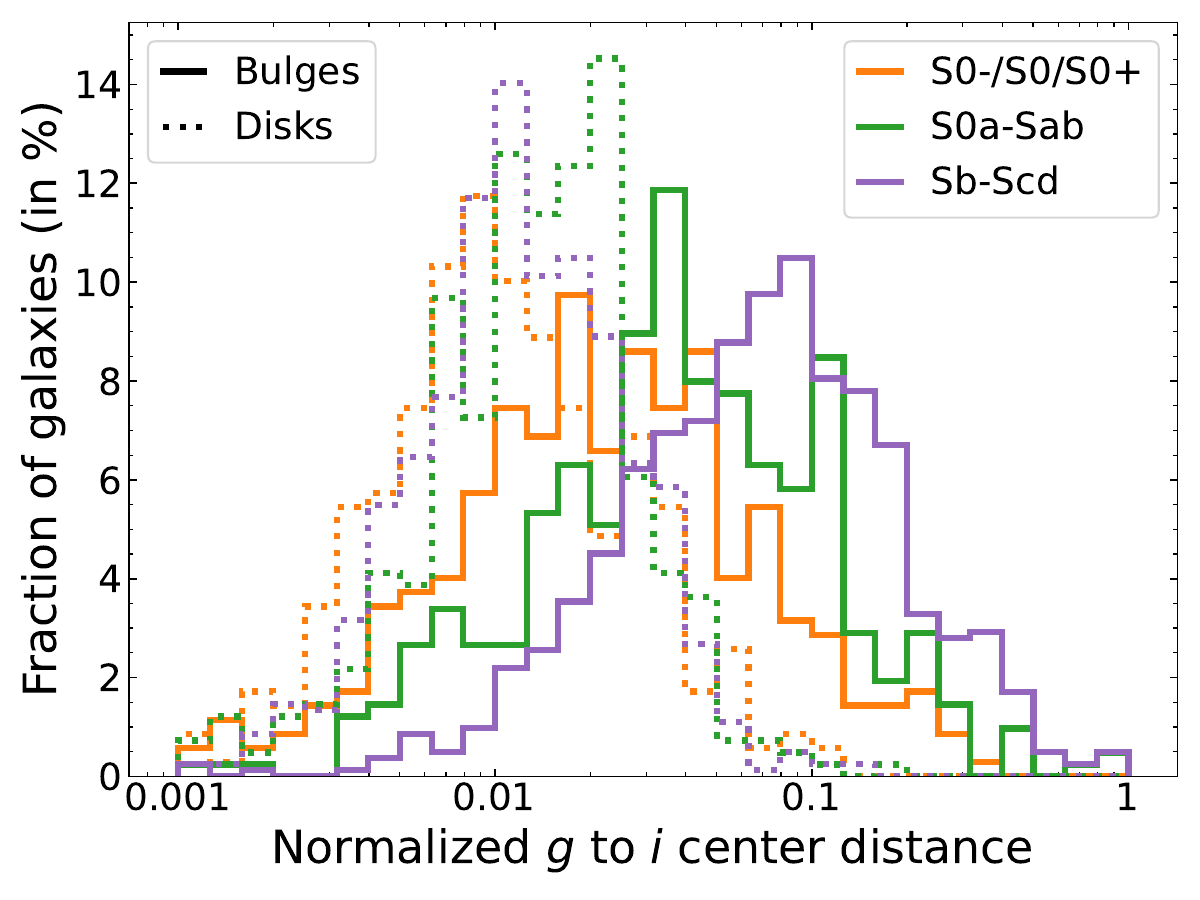}
\caption{Left: Logarithm of the ratio of the effective radii for the $g$ and $i$ bands, versus the distance between the centers of the $g$ and $i$ bulge profiles, normalized by $R_{{\rm e}, g}$ (color coding of the points by type are indicated in the legend). Intermediate-type spirals show larger centering offsets in their profiles in addition to having stronger effective radii variations between bands. Right: Distribution of the distances between the centers of the $g$ and $i$ S\'ersic profiles of the bulges (solid lines) and disks (dotted lines), normalized by their measured effective radii in the $g$ band (the curves are color coded as indicated in the legends). Centering offsets for disks are mostly negligible, whereas for bulges the offsets are overall larger than for disks and get increasingly large for later types.}
\label{fig-center-B-D-dist}
\end{figure*}

To investigate transverse bulge color gradients, we examined centering offsets between all the $g$ and $i$ models normalized by their effective radii. The left panel of \fg\ref{fig-center-B-D-dist} shows centering offsets that are higher and more dispersed for all types of spirals compared to lenticulars together with the stronger $R_{\rm e}$ variation between bands already seen in \fg\ref{fig-color-gradients-bulges}. These offsets in centering range from a few to several tens of percents of the bulge radius. This effect is better illustrated by the distributions of normalized centering offsets for both the bulge and disk models shown in the right panel of \fg\ref{fig-center-B-D-dist} for lenticulars, S0a to Sab types, and Sb to Scd types. The normalized centering offsets are negligible for the disks (dashed lines), with $95\%$ of them being below $0.045$ for S0s, $0.041$ for S0a-Sab, and $0.040$ for Sb-Scd. More noticeable offsets are observed in all bulges (solid lines), with 21.2\% and 8.6\% of S0 types displaying normalized bulge offsets higher than $0.05$ and $0.10$, respectively, and increasingly more so in galaxies of later types, with 39.2\% and 19.4\% of S0a-Sab bulges, and with 63.9\% and 34.6\% of Sb-Scd bulges, having offsets larger than both values, respectively. Moreover, the median normalized offsets are $0.023$, $0.038$, and $0.071$ for the three classes of bulges (from earlier to later types) respectively, compared to $0.010$, $0.014$, and $0.012$ for disks.

We emphasize that among the 85 galaxies in total that were examined visually, we did not see a galaxy bulge with radial gradient in color that would appear as a ring of different color from the bulge interior. However, a significant fraction of the examined bulges do show regions of obvious extinction, sometimes located in the disk at the outskirts of the bulge, which appears to contribute to the bulge profile shifts. Unlike disk gradients that are radial, the bulge color gradients in intermediate-type spirals appear mostly transverse. The increasing differences between bands in the effective radii, S\'ersic indexes, and profile center positions of bulges, as well as the increased dispersion in these parameters, all indicate more complex color profiles for the bulges hosted in spirals of types Sa to Scd, compared to more uniform bulges within the lenticular galaxies. Examining the bulge color gradients as a function of stellar mass also suggests that they are shaped by morphology rather than mass within each morphological type, and the largest differences in gradients are a function of morphological type, not mass (see Appendix~\ref{sct-appendix-mass}).

\section{Comparison with other analyses \label{sct-compare}}

Many recent analyses based on bulge and disk decompositions of large samples of galaxies agree with the EFIGI results that bulges are redder than their host disks \citep{Mollenhoff-2004-BD-spirals-UBVRI, Vika-2014-multiband-BD-decomposition-MegaMorph, Kennedy-2016-GAMA-color-gradients-vs-BT-color-BD, Casura-2022-bulge-disk-decomposition-GAMA}. The $0.09 \pm 0.01$ and the $0.11 \pm 0.02$ average color differences measured by \cite{Head-2014-bulge-disk-colors-ETG-Coma-cluster} for 200 S0 galaxies in the Coma cluster, and by \cite{Barsanti-2021-bulge-disk-colors-clusters} for 469 S0 from eight nearby clusters, respectively, are comparable with the $0.07 \pm 0.01$ median value obtained here for the 105 S0$^+$ galaxies in EFIGI. \cite{Vika-2014-multiband-BD-decomposition-MegaMorph} examined the color difference between bulges and disks as a function of the Hubble type, but with a much smaller sample of 163 nearby galaxies. The measured $(g-i)$ average color differences are $0.03 \pm 0.04$, $0.05 \pm 0.1$, $0.30 \pm 0.07$, $0.28 \pm 0.1$ for their E, S0-Sa, Sb-Sc, Sd-Irr samples, respectively, and indicate an increasing color difference along the Hubble sequence, starting from a compatible-with-zero difference for early-type galaxies and becoming significant for spiral types. This is in agreement with the trend in \fg\ref{fig-color-D-vs-B} and these values are compatible with the EFIGI average color differences of $-0.04 \pm 0.01$, $0.11 \pm 0.01$, and $0.56 \pm 0.01$ for the E, S0-Sa and Sb-Sc samples, except for the stronger difference we find for intermediate-type spirals. An increasing bulge-disk color difference with Hubble type was also found by \cite{Mollenhoff-2004-BD-spirals-UBVRI} from the S\'ersic bulge plus exponential disk decomposition of 26 unbarred bright spirals ($B\le 12.7$) in the $UBVRI$ bands.

For 177 isolated AMIGA galaxies, \cite{Fernandez-Lorenzo-2014-bulges-pseudobulges-relics} found $0.24\pm0.18$ mag and $0.32\pm0.16$ mag offsets between the bulge and disk $(g-i)$ median color for disk-dominated galaxies hosting pseudo-bulges (defined as having a bulge S\'ersic index $n_\mathrm{bulge} < 2.5$), with $B/T < 0.1$ and $0.1<B/T<0.5$, respectively. First, the trend of a larger color difference for larger $B/T$ disagrees with our results, as well as with those of \cite{Mollenhoff-2004-BD-spirals-UBVRI} and \cite{Vika-2014-multiband-BD-decomposition-MegaMorph}. Using the same criterion for $n_\mathrm{bulge}$ and $B/T$ in the EFIGI sample, with the additional exclusion of the lenticular galaxies that are rarely seen in isolation and frequently have  classical bulges, and using $r$-band $B/T$ and $n_\mathrm{bulge}$, we derived offsets of $0.56\pm0.01$ mag and $0.36\pm0.01$ mag, for $B/T < 0.1$ and $0.1<B/T<0.5$, respectively. These color differences are compatible for the high $B/T$ values but diverge for the small $B/T$ sample. We note that \cite{Fernandez-Lorenzo-2014-bulges-pseudobulges-relics} used aperture photometry, which means that the part of the disk profile underlying the bulge region that was subtracted in our bulge-disk decompositions is still present in their study, and may lead to an underestimation of the color difference, with an increasing effect for fainter bulges. Moreover, their minimum aperture radius of 5 pixels leads to the inclusion of disk light in the 15\% smallest bulges. The difference in median bulge and disk color found for low $B/T$ values could then be even more underestimated.

Figure \ref{fig-color-D-vs-B} contradicts the results of \citet{Peletier-1996-bulge-and-disk-colors}, who measured, for early spiral types, color variations from galaxy to galaxy that are larger than the color differences between disks and bulges within galaxies. Figure \ref{fig-color-D-vs-B} also disagrees with \cite{Cameron-2009-MGC-color-concentration-bimod}, who found an overall correlation between the bulge and disk $u-r$ colors. By contrast, \cite{Kennedy-2016-GAMA-color-gradients-vs-BT-color-BD} found that bulges are redder than their disks. However, they also found that bluer galaxies show similar bulge and disk colors, whereas redder galaxies have bulges redder than their disks, contrary to the trend of increasing bulge-disk color difference for bluer galaxies measured here.

\cite{Natali-1992-disk-gradient-NGC-628} decomposed NGC 628 as the sum of a de Vaucouleurs bulge and an exponential disk in $UBVRI$ and were the first to measure a variation in the disk scale-length with wavelength, finding a blue gradient. This result was then extended by using the disks of seven face-on spirals in \citealt{Pompei-Natali-1997-7-spiral-disk-gradients}. Blue disk gradients were also found by \cite{de-Jong-1996-BD-decomposition-IV-stellar-dust-gradients} who performed bulge and disk decomposition of 86 face-on spirals in the $BVRIHK$ bands. Although a detailed comparison with the EFIGI disk gradients is complicated by the different bands used, these various results are analogous. Moreover, \citet{Mollenhoff-2004-BD-spirals-UBVRI} measured decreasing and increasing effective radii of disks and bulges, respectively, across Johnson bands from $U$ to $I$, which may match the color gradients we measured. 

Finally, we note an agreement in the blue disk gradients measured here in EFIGI for the nearby Universe, with those extended up to $z < 0.6$ in \cite{Quilley-2025-ERO}, but these results do not show bulge color gradients. This may be due to their small angular effective radii, especially at larger redshifts, and also to the absence of priors in the modeling, which might be needed to inspect such small-scale effects.
 
\section{\label{sct-discussion}Discussion}

One major result of the present analysis is that bulges of intermediate-type spirals (Sb through Scd types) have the reddest and most dispersed $g-i$ colors, the largest variations in their color gradients (ratio of effective radii in $g$ and $i$) and the largest relative changes in the steepness of their light profiles with color (S\'ersic index $n$). We interpret these systematic changes with Hubble type as resulting from patchy dust differentially extinguishing and scattering the bulge light in the $g$ and $i$ bands. First, dust would generate redder $g-i$ colors, and could explain why intermediate-type spirals have redder bulges than the more metal-rich and older bulges of the early-type galaxies. Indeed, these effects cannot be attributed solely to the ages or metallicities of spiral bulges, as the bulges of elliptical and lenticular galaxies are old and metal-rich, yet are not as red as the bulges of intermediate-type spirals.

Second, a complex distribution in the patchy extinction impacts more the emission in the $g$ band, and hence reduces the $g$ effective radius compared to $i$. Third, it would also lead to a flatter profile in $g$ than in $i$, in order to model the central and external bulge light beyond the regions of patchy dust (see \fg\ref{fig-sersic-color-gradient}). Finally, irregularities in the dust distribution or slight inclination of the disk relative to the line of sight would lead to stronger dust extinction on a side of the bulge compared to the other and may lead to a bluing of the bulge light due to scattering and also to transverse color shifts (see \fg\ref{fig-bulge-exple}), which is measured quantitatively in centering offsets in the multiband models (\fg\ref{fig-center-B-D-dist}). The effect can be seen in some of the bulge images shown in \fg\ref{fig-bulge-exple}.
These images also illustrate the frequent nuclear features such as inner bars, inner spiral arms, rings, or star clusters, which we observe in the residual maps obtained by subtracting the bulge and disk profiles from the data images (see \fg\ref{fig-bulge-exple}), which lead us to hypothesize that these structures also contribute to the detected effects.

The measured disk and bulge gradients and bulge dispersions, as well as the residual structures seen in and around the bulges, depend on the reliability of the bulge-disk decomposition. We emphasize that we did not make overly constraining assumptions in this modeling (e.g., common centers of the bulge and disk models in the various bands), in order to not mask real astrophysical effects. This flexibility as well as careful checks on the choices of priors in the bulge-disk decompositions is what contributed to the detection of transverse color gradients in bulges. We consider that the large dispersion in the disk and bulge gradients are not due the uncertainties or reliability of the profile fitting. They are symptomatic of the complexity of the light distribution of galaxies revealed by this analysis.

Some of the photometric and morphological features that are observed within the disks (bars, spiral arms, and rings) are capable of -- or are signs of -- inward gas flows across the disks. These flows of gas may eventually reach into the nuclear region, leading to regions of dense gas in which star formation can be triggered. We actually see evidence of star formation on the outskirts of some of the bulges, generally in the form of blue knots along inner rings or inner spiral arms. Moreover, the structures we find in the bulges and the dust distribution are related. For both the diffuse warm and cold neutral medium and the cold dense gas, a significant fraction of their masses is contained in the form of dust \citep[see][and references therein]{Eales-2012-gasmass-dustmass, Shull2021-H2-EB-V}. If gas is dynamically swept along spiral arms and bars, for example, it would lead to regions of higher density and high dust opacity within the bulge. There is some theoretical support for this picture of dust influencing the photometric properties and creating the complexity of light distribution in galaxies, especially for bulges. Models of the impact of the distribution of dust extinction that can impact the half-light radii and steepness of the photometric profile of bulges as a function of color follow roughly similar trends \citep[see, e.g., ][]{Pastrav-2013-dust-photometry-disk-bulge-spirals, Pierini-2004-dust-extinction-bulge-disk, Gadotti-2010-dust-extinction-bulge-disk}.  
\citet{Quilley-2022-bimodality} show that in the NUV-$r$ versus stellar mass plane there are three general features – the Blue Cloud, made up of star forming spiral-type galaxies, the Green Valley, comprising early-type spirals (predominately S0a, Sa, and Sab galaxies), and the Red Sequence, which comprises early-type passive galaxies (S0-E types). They also show that there is a continuous bulge growth along the Hubble sequence all throughout this diagram, from the bulge-less irregular and late-type spirals at the low-mass end of the Blue Cloud, to the Sb type, the most massive in the Blue Cloud, with ${B/T}_g \sim 0.15$, and then up to ${B/T}_g \sim 0.45$ for lenticular galaxies, while remaining in the same stellar mass interval.

In the present analysis, the morphological features and dust detected in and around the bulges are interpreted as bringing additional evidence of bulge growth along the Blue Cloud and into the Green Valley. The observation that bulges along this sequence have gas provides some direct evidence of in situ star formation. The dispersion in the colors and gradients indicate that the in situ growth and gas content are stochastic (see also \fg\ref{fig-bulge-gradients-vs-stellar-mass}). Other processes, such as dynamically heating stars within the inner regions of spirals and lenticular galaxies due to bars and inner spirals, can also lead to the growth of the stellar masses of bulges. These critical changes in bulge colors and color gradients, and in their dispersion, occur not only at the entry of the Green Valley but also around the intermediate-type spirals that have bulges that are starting to shift onto the Kormendy relation -- specifically onto the relation between effective surface brightness and magnitude -- which was derived for early-type galaxies \citep[this has been historically used as a criterion to differentiate between pseudo-bulges and classical bulges; see][and references therein]{Quilley-2023-scaling-relations}. The complex color profiles of the bulges of Sb to Scd galaxies are therefore consistent with the fact that many of them could be built via a number of complex secular processes.

The stellar mass of galaxies varies dramatically along the Hubble sequence, as \citet{Quilley-2022-bimodality} showed that it increases by about a factor of 10$^4$ from late-type spirals to intermediate-type spirals (e.g., Sb type galaxies), whereas the difference in the total stellar masses of Sb-type and lenticular galaxies is relatively small. At the same time, the bulge stellar masses increase by about a factor of 10 from Scd to Sb galaxies and then further increase by a factor of 2 from Sb-type galaxies to lenticular galaxies. In addition to these differences in mass, the low-mass galaxies (with masses of $\approx10^9$ M$\sun$) -- that is, the very late-type spirals and irregulars -- have local co-moving volume densities that are approximately a factor of 5 higher \citep{Weaver-2023-COSMOS2020-SMF} than those of the intermediate and lenticular galaxies (with masses of $\approx10^{11}$ M$\sun$).  The low-mass gas-rich spirals must therefore grow through mergers to decrease their co-moving volume densities with time (which rules out that gas accretion alone can drive this evolution). These galaxies must have relatively high gas-to-stellar mass ratios to reform disks and grow along the Blue Cloud. This is interesting, as we have found and others have suggested that even if merging is important for the growth of galaxies, the late and intermediate-type spirals show a similar phenomenology and very similar morphological traits. It is also the increasing relative gas depletion from late-type spirals through lenticular galaxies that drives the morphological change and halts the growth of disks and bulges. Overall, these processes are slow and appear to be key in shaping the color-mass distribution of present-day galaxies.

We caution here that the EFIGI sample is not complete in mass, but we do find that the color gradients of the bulge appear to be shaped by morphology rather than by mass (and it is similar for disk color gradients). Specifically, for a given type, the bulge color gradients do not depend significantly on mass; therefore, a change in their properties in the Green Valley transition region (in color and luminosity-weighted age) appears to be the result of a change in the Hubble type, not a change in the total stellar mass (see Appendix~\ref{sct-appendix-mass}).

As far as disks of spirals are concerned, they have a larger area than bulges and so attributing their smaller half-light radii in $i$ versus $g$ to the impact of clumpy dust is difficult without more information. Dust might play a role, as the strongest color gradients are found in intermediate-type spiral galaxies, which are those with the largest values of the \texttt{Visible Dust} attribute. Nevertheless, we do not find a direct correlation between the amount of visually detectable dust and the color gradients of the spiral and lenticular disks. 

We note that the negative color gradients observed here match the negative age gradient, the negative metallicity gradient, and the decreasing dust extinction with radius, found for intermediate-type spirals by \cite{Gonzalez-Deldago-2015-CALIFA-Hubble-sequence-age-metallicity-gradients}. Within the current study, we cannot disentangle these complex effects on the statistical properties of color gradients within disks, but we suggest some physical mechanisms that would lead to the observed gradients. 

The variations in the disk effective radii between bands measured here could result from smooth and progressive color variation with radius, throughout the extent of the disk, but they could also be the consequence of a discontinuity in the disk color profile. One possible cause of smooth variations with radius could be the different scale lengths of an older thick disk and a younger thin disk \citep[see][and references therein]{Haywood2018MWsummarypaper}. Various discontinuities could be caused by the morphological features within the disks. The frequency of bars is high and spiral disks are defined by having spiral arms. For example, \citet{de-Lapparent-2011-EFIGI-stats} estimated that within the EFIGI sample approximately 30 to 60\% of spiral galaxies (from late to early types) have bars \citep[see also][]{Lee2019barfractions}. Bars, which have finite lengths \citep[and tend to be redder than their disks;][]{Kruk-2018-bar-galaxy-zoo}, rings, and spiral arms could all contribute to discontinuities in the disk color profiles.

Although we have isolated the photometric properties of bulges and disks separately, their evolution is linked. The bars, spiral arms, rings, and lenses that are observed in galaxies are either capable of driving gas inwards or are the result of gas inflows. This inward flow can then support star formation in the inner disks and bulges of galaxies. The gas flows and star formation will eventually deplete the gas, and without gas the defining morphological features of spirals galaxies weaken such that intermediate-type spirals move slowly toward becoming lenticular galaxies. More work is needed to solidify this picture in which mergers and perturbations to the gas due to morphological features are important for this morphological evolution.

\section{Summary, conclusions, and perspectives\label{sct-summary}} 

We have presented a study of 3106 galaxies with visually determined Hubble-types from the EFIGI sample for which we fit one- or two-component S\'ersic profiles to determine the characteristics of their light distributions. Specifically, we focus on measuring and interpreting the colors, half-light radii, and S\'ersic indexes, of bulges and disks as a function of their overall morphology, characterized by their Hubble-type, as well as their bulge-to-total light ratio. From this careful analysis, we find that:

$\bullet$ The $g-i$ colors of galaxy disks and of whole galaxies show a systematic change, becoming increasingly blue from early to late Hubble types, with a $\sim0.6$ mag change across the full Hubble sequence, which is significant compared to the $\sim0.1$ mag scatter per type. The colors of the bulges, however, show the opposite trend, becoming increasingly red, on average, from early to intermediate-type spirals, with the types Sb-Sc having the reddest mean colors but with only a 0.2 mag change between the bluest and reddest bulge type, and hence on the order of the bulge color scatter per type. Bulges in spirals are systematically redder than their disks, but this difference progressively fades from late to early types.

$\bullet$ The reddening of the disks from late to early types along the Hubble sequence is the consequence of the decrease in the star formation rate per unit of stellar mass. The reddening of bulges in intermediate spirals is shown to be the consequence of patchy dust rather than older metal-rich stellar populations. The increase in $g-i$ colors of galaxies from late to early types is the consequence of both the disk reddening and the bulge growth.

$\bullet$ The dispersion in the $g-i$ colors is relatively constant for disks and whole galaxies. However, the colors of bulges have increasing scatter with increasingly later types, with the largest scatter in the colors over a broad range of Hubble types -- from types S0+ through Scd.

$\bullet$ Disks appear on average systematically more compact in the $i$ band than in the $g$ band as one moves from early- (cD through S0a) to later-type galaxies (Sa through Sd), with ellipticals and lenticular galaxies having similar effective radii in the $g$ and $i$ bands. Disks of later-type spiral galaxies, up to Sbc, have systematically smaller half-light radii in the $i$ band compared to the $g$ band. The very late-type spiral disks, Sd to Sdm, show similar median effective radii in both the $g$ and $i$ bands. 

$\bullet$ Both the highest values and dispersions in the bulges, $R_{\mathrm{e},g}/R_{\mathrm{e},i}$, occur for the same intermediate-type (Sa-Sc) spiral galaxies that have the highest dispersion (by a factor of 2 to 3) in bulge colors and the reddest colors. These galaxies also show the largest differences in their $g$ and $i-$band S\'ersic indices, and the largest distances in the centering of the $g$ and $i-$ profile models, which all result from transverse color gradients in their bulges. The larger effective radii and steeper profiles of the bulge in $i$ compared to the $g$ band for intermediate-type spirals, as well as the spatial offsets in their centers, all provide further evidence that patchy dust is shaping the light profiles of these bulges. 

$\bullet$ Moreover, in most of the residual images of the bulges that we have examined, we see evidence of nuclear substructures such as bars (often associated with a larger bar in the disk), inner spiral arms, circumnuclear rings, and nuclear star clusters (clumpy regions that appear blue or are relatively bright, which we interpret as star-forming clusters).

$\bullet$ For a given type, the bulge color gradients do not depend significantly on stellar mass. The change in their properties as one moves from the end of the Blue Cloud through the Green Valley to the Red Sequence in the mass versus color plane appears to be the result of a change in the Hubble type, not a change in the total stellar mass.

These results fit within a picture of galaxies evolving (backward) along the Hubble sequence from late to earlier types, as is suggested in \cite{Quilley-2022-bimodality}. Late-type galaxies have the bluest colors and do not have significant color gradients, and hence have similar star formation histories as a function of radius. Then, intermediate spirals, with higher masses and small bulges ($B/T \leq 0.15$), have complex bulge color profiles and the reddest bulges, due to patchy dust. Using the dust as a surrogate for gas indicates, along with some morphological structures such as bars and inner spirals, that some of the bulge growth occurs due to star formation. The morphological structures also indicate that some bulge growth occurs through the dynamical heating of stars in the stellar disk \citep[secular evolution leading to the creation of pseudo-bulges; see, e.g.,][and references therein]{Quilley-2023-scaling-relations}. The transition from Sb to S0 corresponds to the full transition throughout the Green Valley of the color versus stellar-mass plane, and is characterized by a concomitant bulge growth and shutdown of the star formation in disks (as is seen from their increasingly redder colors). Since we see little evidence of patchy dust in lenticulars, during the transition across the Green Valley, there is also a decrease in the quantity of gas overall (in agreement with direct determinations of the gas content of early-type galaxies). 

Of course, the present analysis only provides circumstantial evidence of in situ bulge growth and what causes the systematic changes in the half-light radii of the bulges and disks across bands. More detailed studies linking bulge growth, gas, and morphological features of bulges and disks are required. The present comparison of the color profiles of galaxies with aspects of their morphology such as bars and rings will give us a further insight into whether these morphological features drive systematically galaxies along an evolutionary sequence or whether they are simply a consequence of isolated dynamical events.

To further our understanding of what drives the systematic changes in the galaxy population will require detailed studies at longer wavelengths and higher resolution of a large sample of galaxies with well-determined Hubble types. The wide-field surveys with Euclid, the Vera Rubin Observatory, and the Nancy Grace Roman space telescope will provide the wide wavelength coverage, depth, and spatial resolution necessary to investigate the morphological properties in greater detail. JWST and the deep surveys with Euclid and Roman telescopes allow us to extend these analyses to higher redshifts and to add information to link the characteristics of galaxies in the local Universe with those at higher redshifts \citep[see, e.g.,][]{Yu-2025-color-gradients-disk-JWST-z-1-3, Jin-2024-color-gradients-z-4-central-SF}.

\begin{acknowledgements}

Louis Quilley acknowledges funding from the CNES postdoctoral fellowship program. Matthew Lehnert is an associated researcher at the Institute d'Astrophyique de Paris and thanks the staff for their support.
This research made use of {\tt SourceXtractor++}, an open source software package developed for the Euclid satellite project. This research made use of the VizieR catalog access tool, CDS, Strasbourg, France (DOI: 10.26093/cds/vizier). The original description of the VizieR service was published in A$\&$AS 143, 23. This research also made use of NASA's Astrophysics Data System.
This work is based on observations made with the NASA Galaxy Evolution Explorer. GALEX is operated for NASA by the California Institute of Technology under NASA contract NAS5-98034. Funding for the SDSS and SDSS-II has been provided by the Alfred P. Sloan Foundation, the Participating Institutions, the National Science Foundation, the U.S. Department of Energy, the National Aeronautics and Space Administration, the Japanese Monbukagakusho, the Max Planck Society, and the Higher Education Funding Council for England. The SDSS Web Site is http://www.sdss.org/.
The SDSS is managed by the Astrophysical Research Consortium for the Participating Institutions. The Participating Institutions are the American Museum of Natural History, Astrophysical Institute Potsdam, University of Basel, University of Cambridge, Case Western Reserve University, University of Chicago, Drexel University, Fermilab, the Institute for Advanced Study, the Japan Participation Group, Johns Hopkins University, the Joint Institute for Nuclear Astrophysics, the Kavli Institute for Particle Astrophysics and Cosmology, the Korean Scientist Group, the Chinese Academy of Sciences (LAMOST), Los Alamos National Laboratory, the Max-Planck-Institute for Astronomy (MPIA), the Max-Planck-Institute for Astrophysics (MPA), New Mexico State University, Ohio State University, University of Pittsburgh, University of Portsmouth, Princeton University, the United States Naval Observatory, and the University of Washington.

\end{acknowledgements}

\bibliography{biblio.bib}

\begin{appendix}

\section{$B/T$ variations with observing band\label{sct-appendix-BT-fits}}

\begin{figure}[ht]
\centering
\includegraphics[width=\columnwidth]{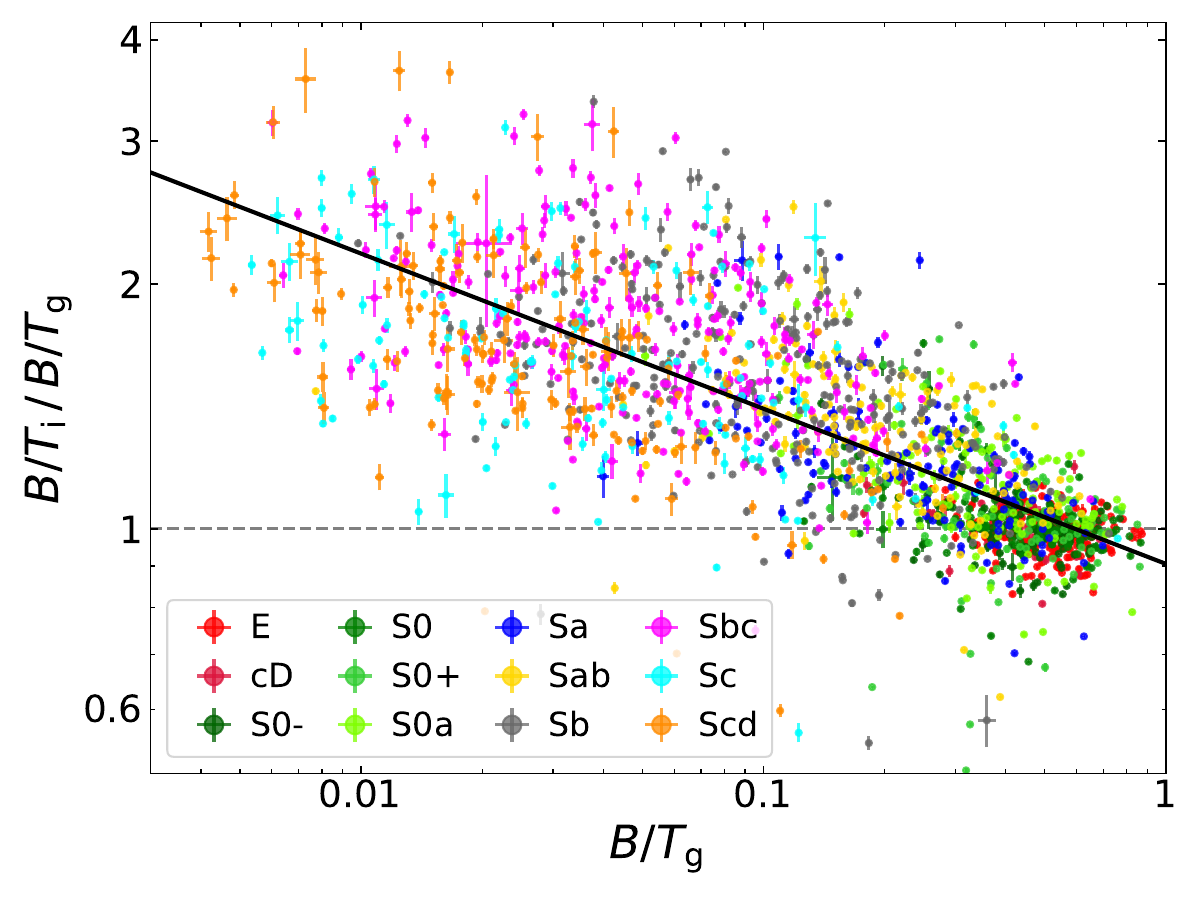}
\caption{Logarithm of the ratio between the bulge-to-total ratios measured in the $i$ and $g$ bands respectively, as a function of ${B/T}_g$, color coded by the Hubble type of the galaxies. The correspondence between colors, point styles, and Hubble types are as indicated in the legend at the bottom left corner of the plot. The black solid line is a linear fit to the data. The bulge-disk color difference leads to a variation of $B/T$ measured across bands, with higher values in the redder $i$ band, up to three times the $g$ band value for $B/T\approx0.01$. This difference steadily decreases towards earlier types in which the $B/T$ values in both bands are similar for the E and lenticular galaxies, whose bulge and disk components have similar colors (\fg\ref{fig-color-B-D}).}
\label{fig-BT-var-gi}
\end{figure}

The differences in color between bulges and disk as a function of morphological type (\fg\ref{fig-color-B-D}), leads to a variation of the bulge-to-total light ratio $B/T$ depending on the band in which it is measured. Higher $B/T$ are measured in the redder bands. Figure \ref{fig-BT-var-gi} shows that intermediate- and late-type spiral galaxies have the largest bulge-disk color difference (see also \fg\ref{fig-color-B-D}), the ratio of $B/T$ between the $i$ and $g$ bands is the largest, and reaches a factor of $\approx2$ for the latest galaxy types which have the smallest contribution of the bulge to the total galaxy luminosity. The early-type galaxies have bulges that become more prominent and the bulge-disk color-dichotomy fades, and thus the $B/T$ values measured in the $g$ and $i$ bands converge to similar values. We perform a linear fit of the data points (\fg\ref{fig-color-B-D}) which is justified by the Pearson correlation coefficient of $r=-0.75$, and obtain the following relations:
\begin{equation}
    \log(\frac{{B/T}_i}{{B/T}_g}) = (-0.191 \pm 0.004)\log({B/T}_g) + (-0.043 + 0.004)
    \label{eq-ratio-BT-gi-vs-BTg}
,\end{equation}

or equivalently that
\begin{equation} 
    \log({B/T}_i) = (0.809 \pm 0.004)\log({B/T}_g) + (-0.043 + 0.004)
    \label{eq-BTi-vs-BTg}
,\end{equation}
with a $R^2$ score of 0.57 and a \rms scatter around the fit of 0.088. This relation can, for instance, be used to express the color variations with ${B/T}_i$ instead of ${B/T}_g$ (see \fg\ref{fig-color-B-D-vs-BT}, \eqs\ref{eq-gi-bulge-color-BT}, \ref{eq-gi-disk-color-BT}, and \ref{eq-gi-galaxy-color-BT}).

Very similar results are obtained for the ratio between the $r$ and $g$-band $B/T$ but with smaller highest values at low ${B/T}_g$, as well as a flatter slope. Lenticular galaxies have similar $B/T$ values in each band. As a function of ${B/T}_g$, this ratio has a Pearson correlation coefficient of $r=-0.69$ which justifies making a linear fit to the data. Doing this, we obtain
\begin{equation} 
    \log({B/T}_r) = (0.875 \pm 0.003)\log({B/T}_g) + (-0.015 + 0.003)
    \label{eq-BTr-vs-BTg}
,\end{equation}
with an $R^2$ score of 0.47 and a \rms scatter around the fit of 0.071.

\section{Bulge-disk color differences\label{sct-appendix-BD-color-diff}}

\begin{figure*}[ht]
\centering
\includegraphics[width=\columnwidth]{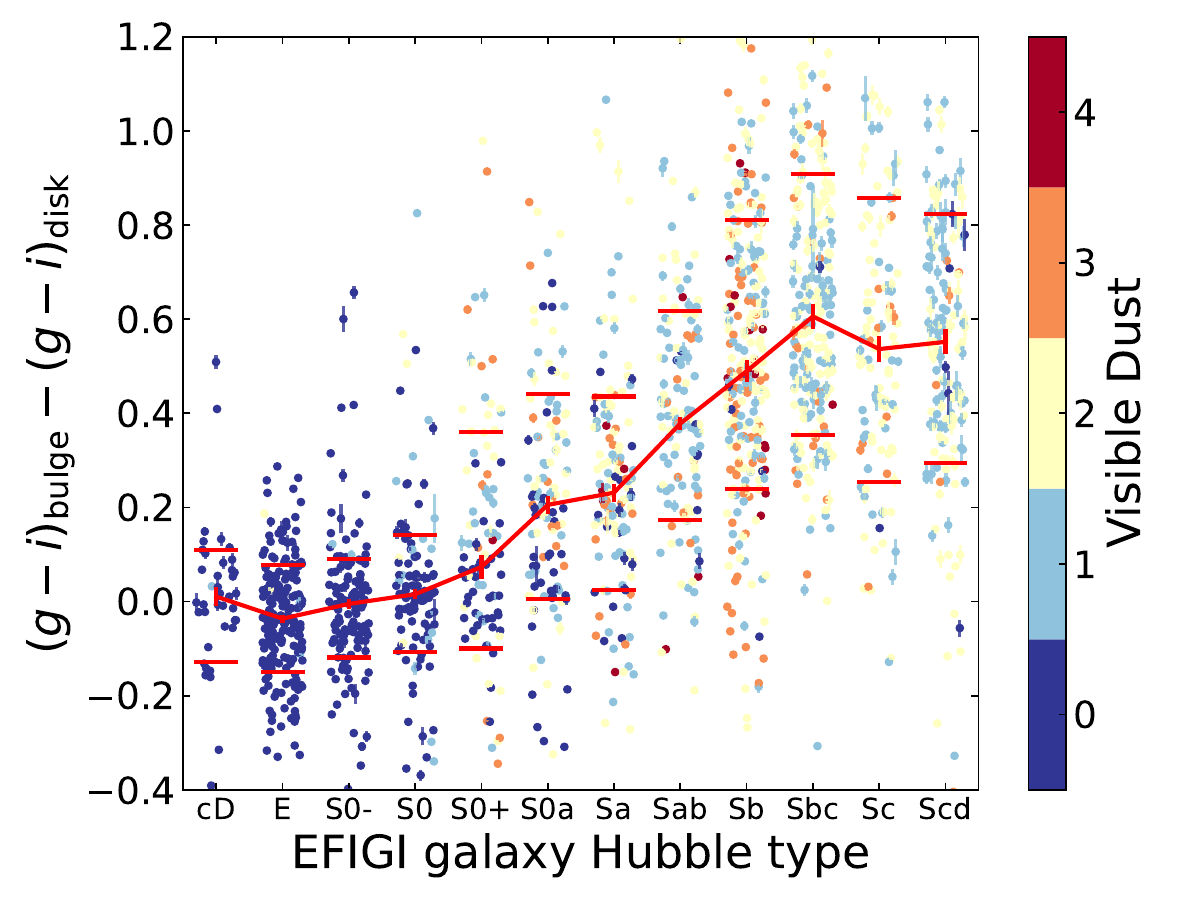}
\includegraphics[width=\columnwidth]{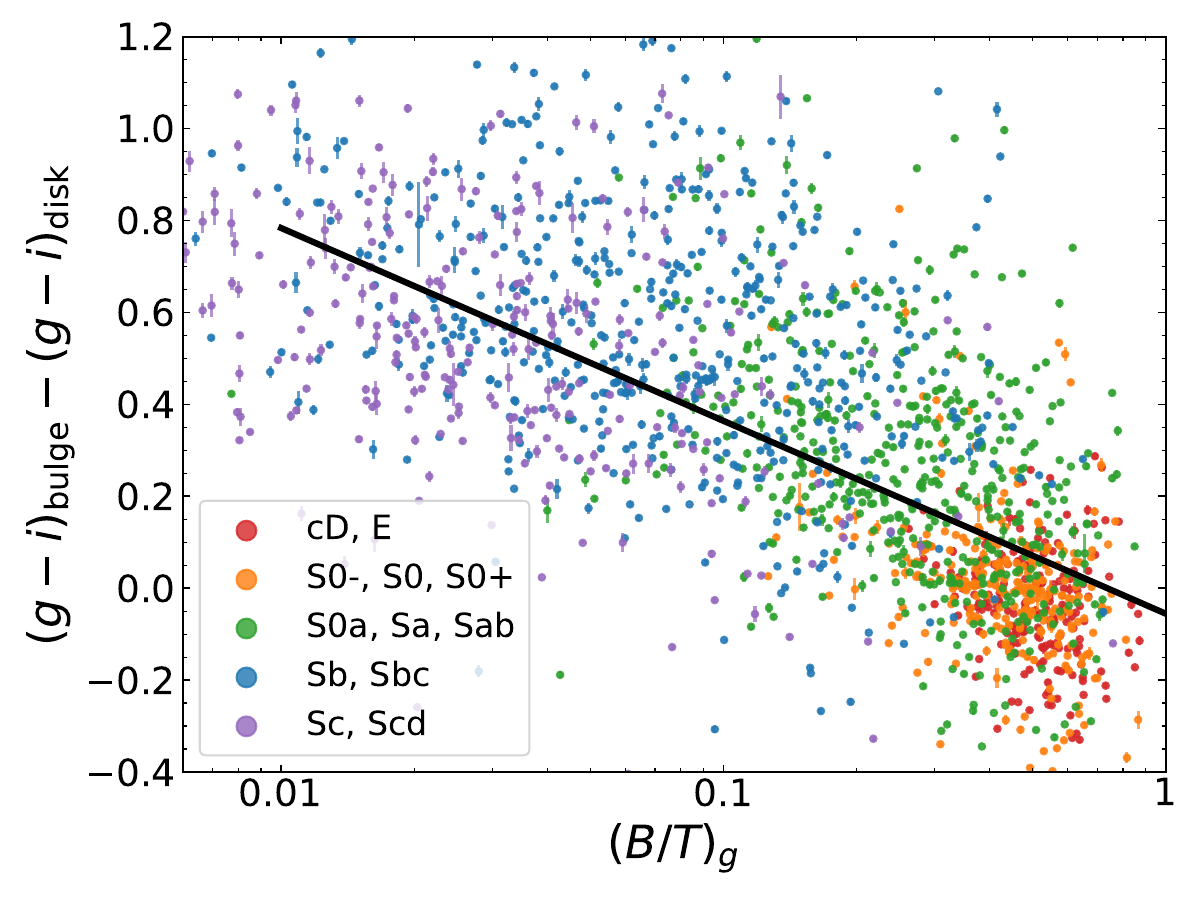}
\caption{Difference between the bulge and disk $g-i$ color for all EFIGI galaxies with \texttt{Incl-Elong}$\leq 2$, as a function of Hubble type in the left panel, and as a function of ${B/T}_g$ in the right panel. The points in the panel on the left are color coded by the \texttt{Visible Dust} attribute as indicated in the legend to the right of the panel. The points in the right panel are color coded by Hubble type as indicated in legend in the bottom right corner of the panel. As already seen in \fgs\ref{fig-color-B-D} and \ref{fig-color-B-D-vs-BT}, bulges and disks are similarly red for early-type galaxies, and disks get progressively bluer than their bulge towards later-type galaxies and smaller $B/T$ values.} 
\label{fig-color-diff-BD}
\end{figure*}

Figures \ref{fig-color-B-D} and \ref{fig-color-B-D-vs-BT} allowed us to discuss the difference in color between bulges and disks in galaxies by comparing their respective distributions. Here, \fg\ref{fig-color-diff-BD} shows directly the distribution of the difference between the bulge and the disk $g-i$ color for each EFIGI galaxy as a function of Hubble type (left panel), and as a function of $B/T$ (right panel). The left panel shows that ellipticals and S0s have compatible with zero median bulge-disk color differences of $-0.04$, $-0.01$, $0.02$ for types E, S0$^-$ and S0, but this difference increases with increasingly later Hubble types, to reach a peak value of $0.61$ mag for Sbc galaxies, then plateaus until Scd types. The various statistics regarding the bulge-disk color difference are given in Tables~\ref{tab-colors-gi} and \ref{tab-colors-gr}.

The right panel of \fg\ref{fig-color-diff-BD} shows how the $g-i$ bulge-disk color decrease with $B/T$ for Scd and earlier types. The linear fit of $(g-i)_\mathrm{bulge} - (g-i)_\mathrm{disk}$ as a function of $\log({B/T}_g)$ is justified by the large negative Pearson correlation coefficient of -0.61 between these two variables, yields:
\begin{equation}
    (g-i)_\mathrm{bulge} - (g-i)_\mathrm{disk} = -0.42 \pm 0.01 \log({B/T}_g) + -0.06 \pm 0.01
    \label{eq-gi-bulge-disk-color-diff-BT}
\end{equation}
with a $R^2$ score of 0.37, and a \rms dispersion of 0.30 around the fit. The resulting slope is equal to the difference of those in \eqs\ref{eq-gi-bulge-color-BT} and \ref{eq-gi-disk-color-BT}, as both the reddening of the disk and bluing of the bulge for increasing ${B/T}_g$ reinforce each other. But the fact that the Pearson correlation coefficient and $R^2$ scores are similar to those for $(g-i)_\mathrm{disk}$, whereas the \rms dispersion is similar to the large dispersion around the bulge color fit with ${B/T}_g$, indicates that the disk color variations drive this variation in the bulge-disk color difference. Equation~\ref{eq-gi-bulge-disk-color-diff-BT} can be rewritten as a function of $B/T$ in the $r$ or $i$ bands using \eqs\ref{eq-BTi-vs-BTg} and \ref{eq-BTr-vs-BTg} (see also \fg\ref{fig-BT-var-gi}).

\section{Disk and bulge colors gradients versus $B/T$\label{sct-appendix-gradients-vs-BT}}

\begin{figure*}[ht]
\centering
\includegraphics[width=\columnwidth]{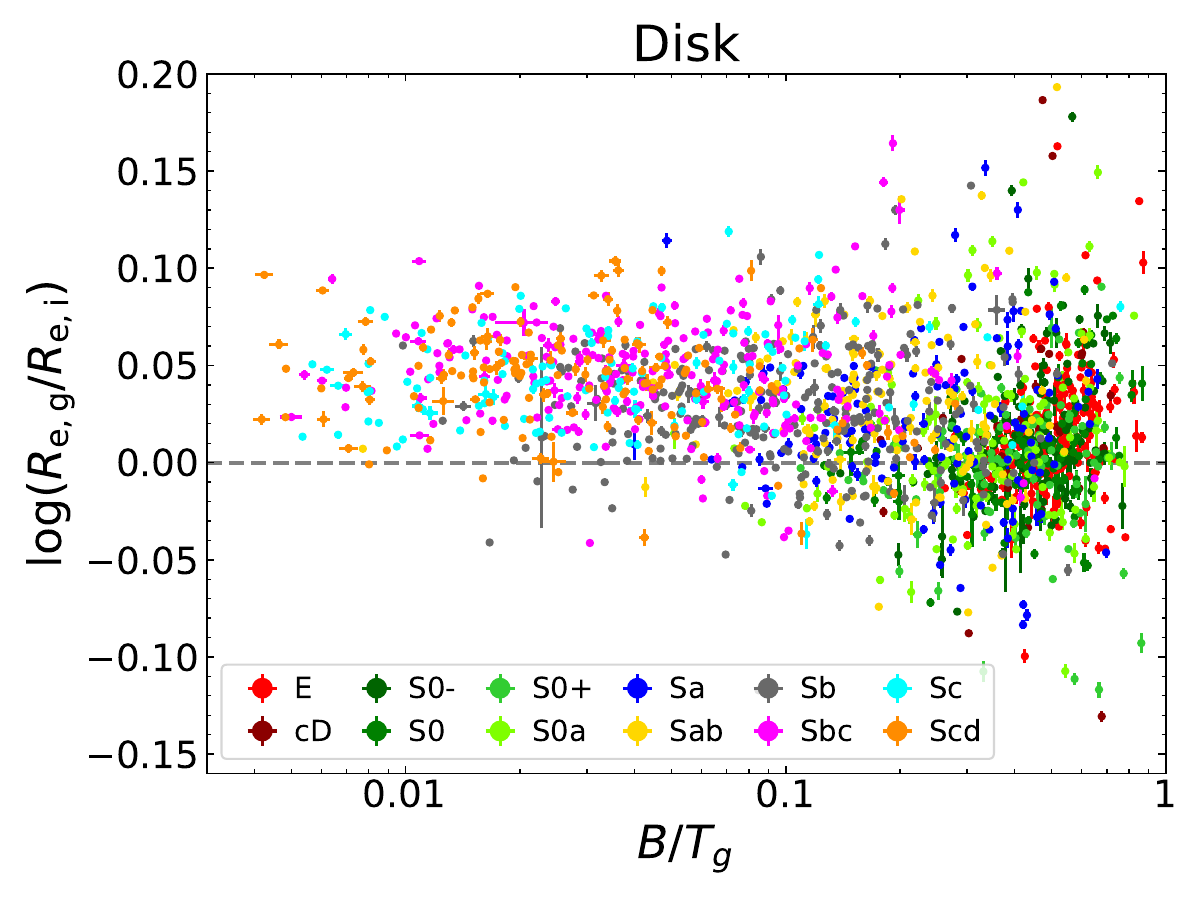}
\includegraphics[width=\columnwidth]{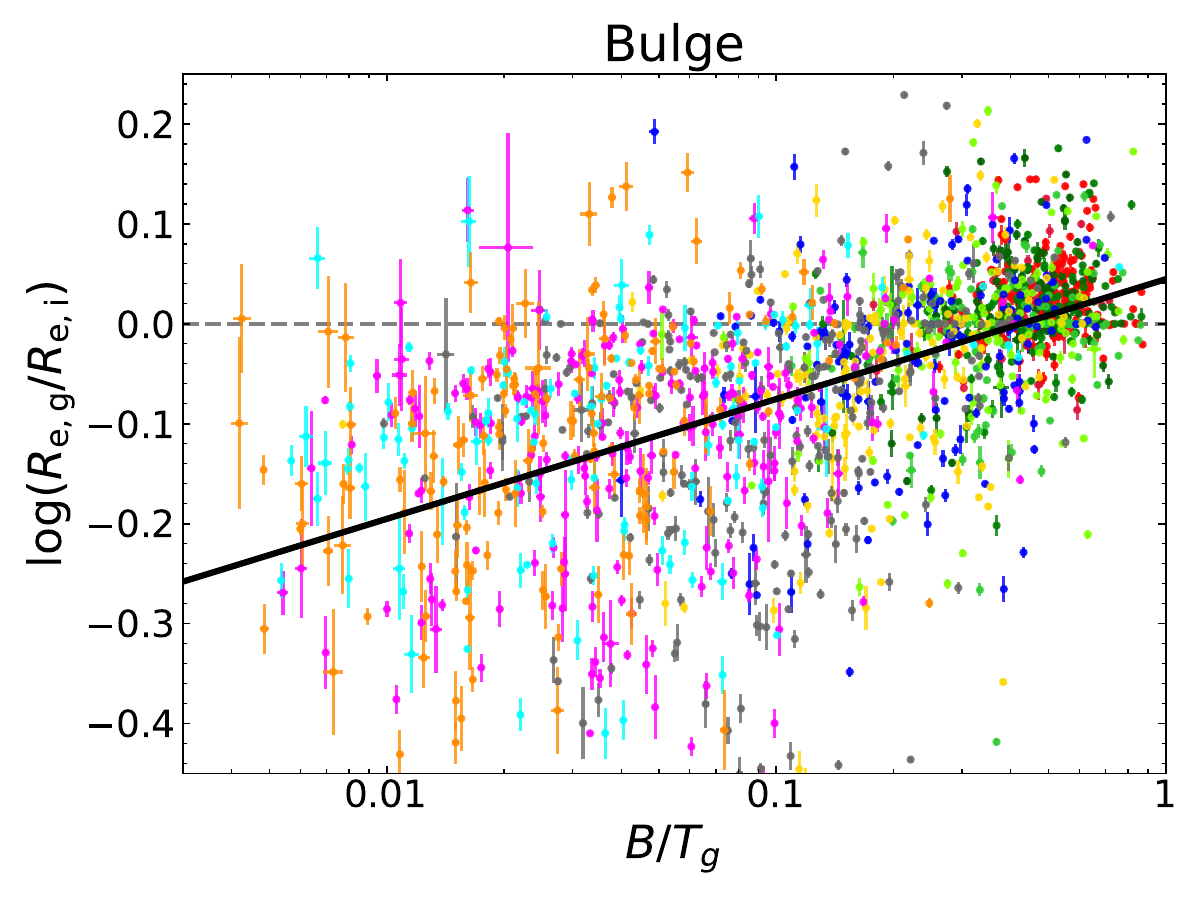}
\caption{The relative distribution of effective radii, shown as $\log(R_{\mathrm{e},g}/R_{\mathrm{e},i})$, for the bulge (left) and disk (right) components of the bulge-disk decompositions of all EFIGI morphological types with \texttt{Incl-Elong}$\leq 2$ as a function of ${B/T}_g$. In both panels, the points are color-coded according to their Hubble type as indicated in the legend in the bottom left corner of the panel on the left. The solid black line in the right panel is a linear fit to the data (see text). Intermediate spirals with ${B/T}_g \leq 0.1$ exhibit the strongest $R_{\rm e}$ variations with bands, in both their bulges and disk, with opposite signs, and with bulges radii having generally stronger variations. At higher ${B/T}_g$ values, the color gradients fade with similar bulge and disk $R_{\rm e}$ across bands.}
\label{fig-color-gradients-BT}
\end{figure*}

Because $B/T$ can be used to parameterize the Hubble type \citep{Quilley-2022-bimodality}, we also examine both the bulge and the disk color gradients as a function of ${B/T}_g$ in the left and right panels of \fg\ref{fig-color-gradients-BT}, respectively. As expected from the variations with Hubble type, on the one hand, for high values of ${B/T}_g$, the disk color gradients of lenticulars and early-type spirals are distributed around 0, with a median value of $\log(R_{\mathrm{e},g}/R_{\mathrm{e},i}) = 0.010 \pm 0.001$ for ${B/T}_g > 0.3$. They increase for lower values of ${B/T}_g$ and are generally positive for ${B/T}_g < 0.1$, which corresponds to intermediate- to late-type spirals, with $\log(R_{\mathrm{e},g}/R_{\mathrm{e},i})$ in the $[0;0.1]$ interval and a median of $0.040 \pm 0.001$ (corresponding to a disk 10\% larger in $g$ than in $i$). In the $0.01$ to $0.1$ ${B/T}_g$ interval, the \rms dispersion in the logarithm of the radii is bounded by 0.022 to 0.028 dex, against 0.035 to 0.042 dex for ${B/T}_g > 0.1$. On the other hand, the median bulge gradients are also close to 0 with $\log(R_{\mathrm{e},g}/R_{\mathrm{e},i}) = 0.013 \pm 0.002$ for large ${B/T}_g$ values in the $[0.3;1.0]$ interval. They decrease towards lower ${B/T}_g$ and are systematically negative for ${B/T}_g < 0.1$, with a median of $\log(R_{\mathrm{e},g}/R_{\mathrm{e},i}) = -0.100 \pm 0.004$. Therefore, the difference between the $g$ and $i$ effective radii is stronger for the bulge than for the disk profiles, as the mean gradient over the same ${B/T}_g$ interval is almost 3 times stronger (and opposite). 

The color gradients in the bulges of spiral types are also more dispersed than those in their disks (similar to the distributions with morphological type, see \scts\ref{sct-disk-gradient} and \ref{sct-bulge-gradient}), with the bulge $\log(R_{\mathrm{e},g}/R_{\mathrm{e},i})$ spread over the $[-0.4;0]$ interval for ${B/T}_g < 0.1$ (the $i$ radius can thus be up to 2.5 times larger than in $g$), and an \rms dispersion bounded by 0.12 to 0.15 dex, contrasting with the lower dispersion between 0.05 and 0.09 dex for ${B/T}_g > 0.2$. This results in a weak correlation as estimated by a Pearson correlation coefficient of $r=0.49$. By fitting a linear relation between the bulge $\log(R_{\mathrm{e},g}/R_{\mathrm{e},i})$ and ${B/T}_g$, we obtain
\begin{equation}
    \log(R_{\mathrm{e},g}/R_{\mathrm{e},i}) = (0.120 \pm 0.005) \log({B/T}_g) + (0.045 \pm 0.005)
    \label{eq-Re-BTg}
,\end{equation}
with an $R^2$ score of 0.24, and an \rms dispersion around the fit of 0.11. Equation~\ref{eq-Re-BTg} can be rewritten as a function of $B/T$ in the $r$ or $i$ bands using \eqs\ref{eq-BTi-vs-BTg} and \ref{eq-BTr-vs-BTg} (see also \fg\ref{fig-BT-var-gi}).

\section{No covariance of bulge and disk color gradients\label{sct-appendix-covariance-BD-gradients}}   

\begin{figure}[ht]
\centering
\includegraphics[width=\columnwidth]{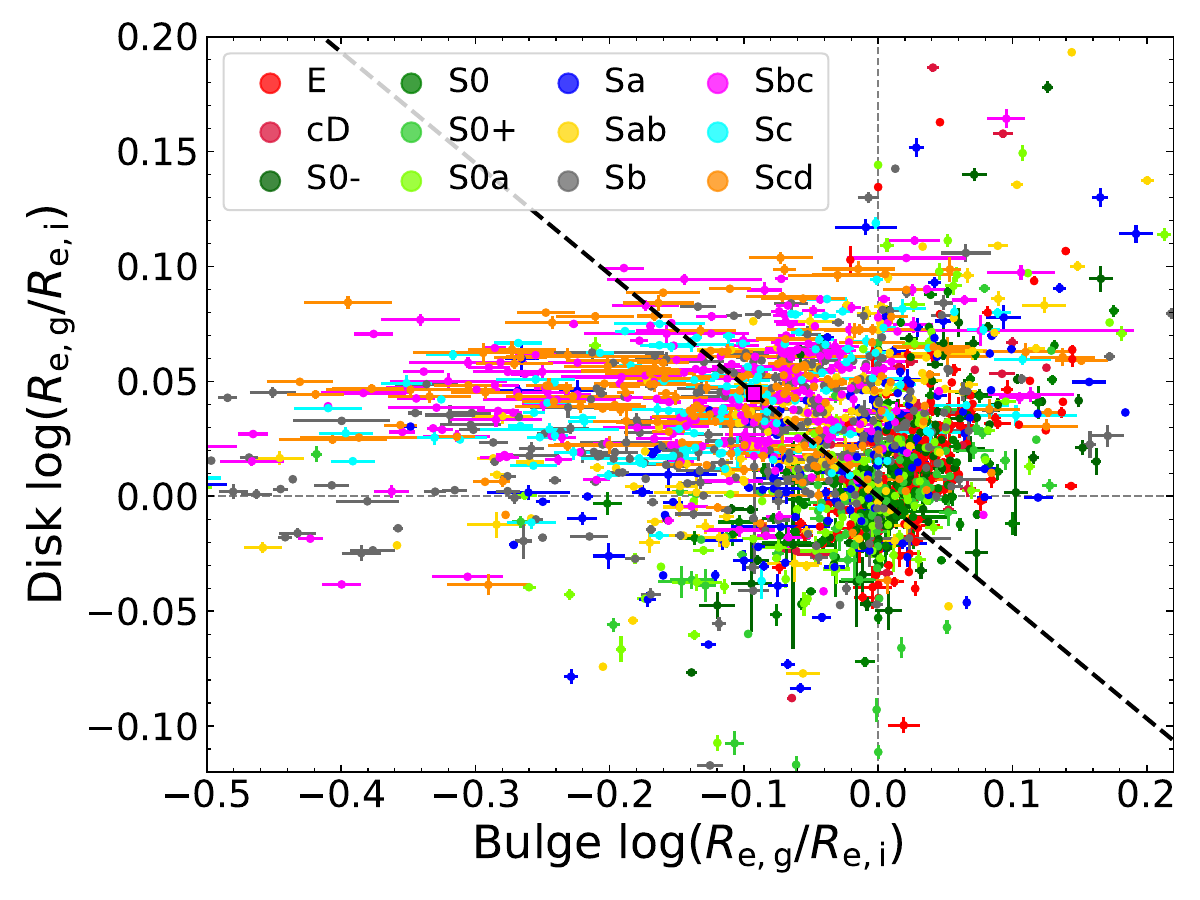}
\caption{Relative disk color gradient versus the relative bulge color gradient for all EFIGI Hubble types (as indicated in the legend). The pink square show the median gradients for Sbc types, and the dashed line correspond to the linear fit constrained to go through the point at which the bulge and disk gradients are both zero. Bulge and disk gradients do not preferentially lie along this a diagonal line (or any diagonal line for that matter), nor are the point distributed perpendicular to it. The two color gradients are completely uncorrelated.}
\label{fig-D-vs-B-gradients}
\end{figure}

The bulge and disk model fits are interdependent in bulge and disk decompositions. To alleviate concerns about the symmetry in the trends actually arising from possible inter-dependencies of fitting two component models to the light distribution of galaxies, rather than what we observe representing true physical effects, 
we directly compare the bulge and disk gradients for all individual galaxies (\fg\ref{fig-D-vs-B-gradients}). 
Instead we see that E and S0 types are spread around the origin, whereas intermediate-type spiral galaxies are spread out over the $\log(R_{\mathrm{e},g}/R_{\mathrm{e},i}) \in [0;0.1]$ interval at all values of the bulge gradients. The dashed line represents the linear relation interpolated from the mean bulge and disk gradients of the Sbc-type galaxies (which displays some of the strongest variations of $R_\mathrm{e}$ with wavelength for both bulges and disks). Data points should lie along this line if the bulge gradient was a consequence of an interdependence between the fits to the bulge and to the disk, which is not the case. This is quantitatively confirmed by computing the Pearson correlation factor between both gradients for each Hubble type, as they all verify $r^2 < 0.35$, indicating a lack of any significant correlation.

\section{Bulge S\'ersic index variations with band\label{sct-appendix-nsersic}}

\begin{figure}[ht]
\centering
\includegraphics[width=\columnwidth]{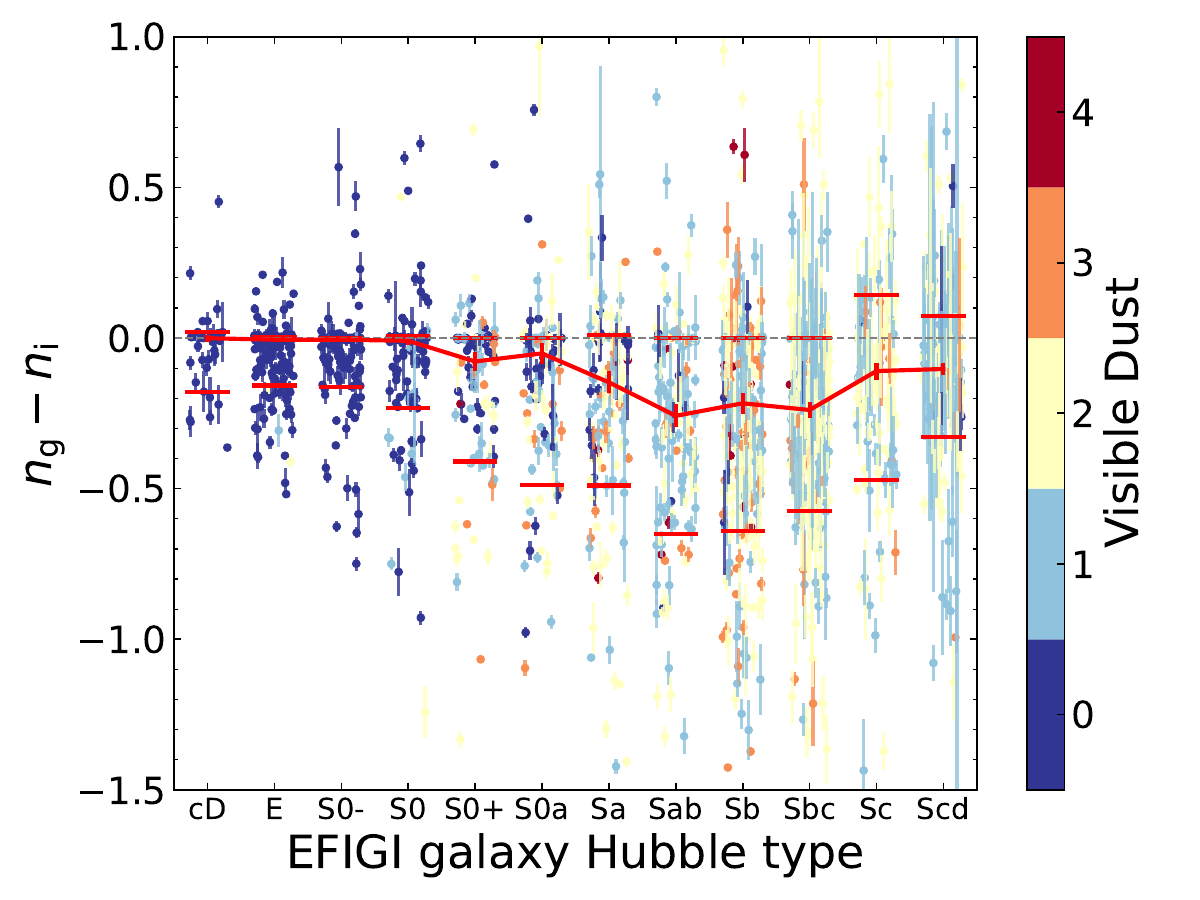}
\caption{Distribution of the difference between the bulge S\'ersic indexes fit in the $g$ and $i$ band as a function of Hubble type. The points are color-coded by the EFIGI \texttt{Visible Dust} attribute. The red dashed lines indicate the median value per type, and the dashed horizontal segments show the 16th and 84th percentiles of each type distribution. A similar trend is found as for the effective radii (see \fg\ref{fig-color-gradients-bulges}), with a small variation in S\'ersic index with wavelength for early-type galaxies and an increasing difference towards later types, which peaks for intermediate spirals, mainly Sab, Sb and Sbc types. For these types, the bulge profile in the $i$-band is steeper than in the $g$-band.}
\label{fig-sersic-color-gradient}
\end{figure}

Figure \ref{fig-sersic-color-gradient} shows the difference between the $g$ and $i$ bulge S\'ersic indexes obtained through the multiband light-profile fitting. The median index differences are consistent with zero for cD to S0 type galaxies, whereas they are negative for S0$^+$ to Scd types (see \tab\ref{tab-nsersic-gradients-gi}). The Sab type galaxies have the lowest median difference of $-0.26\pm0.04$ (using bootstrap uncertainties in the median) and similar values for Sb and Sbc type galaxies. The bulge components of the latest lenticular types and spiral galaxies up to Scd type galaxies have steeper $g-$band profiles than they do in the $i-$band. There is moreover a large dispersion in this index difference for the same types as those for which it is the largest, and this corresponds to galaxies with larger value of the \texttt{VisibleDust} attribute. This is discussed in \sct\ref{sct-bulge-gradient-dispersion}. 

\begin{table*}[ht]
\caption{Statistics of the bulge S\'ersic index $n_g$ and of the difference $n_g - n_i$ between the bulge S\'ersic indexes fit in the $g$ and $i$ bands for all EFIGI morphological types up to and including the Scd type galaxies.}
\resizebox{\linewidth}{!}{%
\begin{tabular}{  l r r r r r r r r r r r r r r r r }
\hline
\hline
\multicolumn{12}{c}{Bulge}\\
\hline
Type & cD & E & S0$^-$ & S0 & S0$^+$ & S0a & Sa & Sab & Sb & Sbc & Sc & Scd  \\
\hline
$N^{(a)}$ & 39 & 203 & 131 & 112 & 105 & 125 & 143 & 144 & 278 & 249 & 124 & 167 \\
Median $n_g$ & 2.646 & 3.003 & 2.594 & 2.728 & 2.488 & 2.387 & 2.391 & 1.875 & 1.776 & 1.496 & 1.353 & 1.049 \\
Error  & 0.187 & 0.061 & 0.073 & 0.115 & 0.094 & 0.117 & 0.086 & 0.071 & 0.079 & 0.089 & 0.118 & 0.074 \\
Median $n_g-n_i$ & -0.001 & -0.006 & -0.006 & -0.009 & -0.078 & -0.051 & -0.147 & -0.258 & -0.217 & -0.239 & -0.110 & -0.103 \\
Error$^{(b)}$ & 0.012 & 0.002 & 0.008 & 0.012 & 0.032 & 0.036 & 0.038 & 0.039 & 0.033 & 0.028 & 0.029 & 0.019 \\
${q_{16}}^{(c)}$ & -0.180 & -0.157 & -0.162 & -0.232 & -0.410 & -0.489 & -0.490 & -0.652 & -0.640 & -0.574 & -0.470 & -0.329 \\
${q_{84}}^{(c)}$ & 0.019 & 0.004 & 0.001 & 0.006 & 0.001 & 0.000 & 0.009 & 0.001 & 0.001 & 0.001 & 0.142 & 0.072 \\
Mean & -0.038 & -0.051 & -0.064 & -0.085 & -0.164 & -0.185 & -0.214 & -0.318 & -0.288 & -0.280 & -0.138 & -0.121 \\
$\mathrm{SD}^{(d)}$ & 0.012 & 0.002 & 0.008 & 0.012 & 0.032 & 0.036 & 0.038 & 0.039 & 0.033 & 0.028 & 0.029 & 0.019 \\
\hline
\end{tabular}
}
\small\textbf{Notes.}
$^{(a)}$ Number of galaxies per Hubble type in which both the bulge and the disk fits (for Scd type galaxies and earlier). 
$^{(b)}$ Error in the median per Hubble type, determined by bootstrapping.
$^{(c)}$ 16\% and 84\% percentiles of the distribution per Hubble type.
$^{(d)}$ Standard deviation of the distribution per Hubble type.
\label{tab-nsersic-gradients-gi}
\end{table*}

\section{No significant impact of bars on bulge colors and color gradients \label{sct-appendix-bar}}

\begin{figure*}[ht]
\centering
\includegraphics[width=\columnwidth]{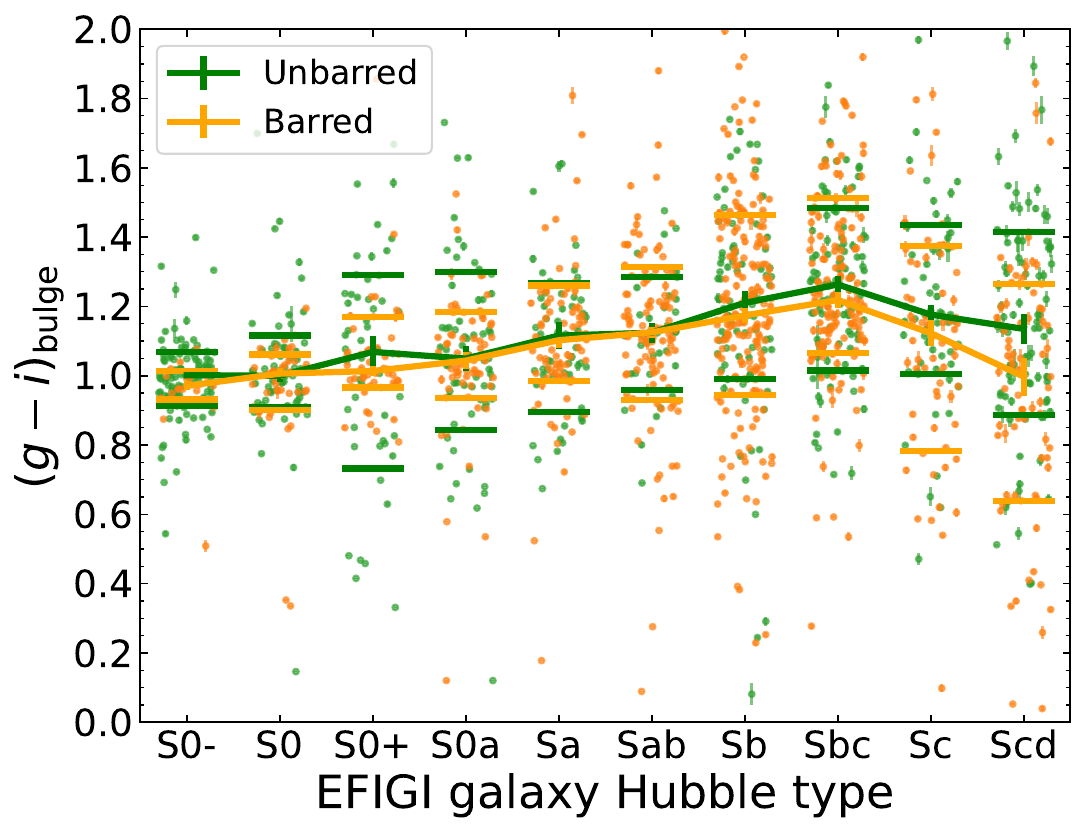}
\includegraphics[width=\columnwidth]{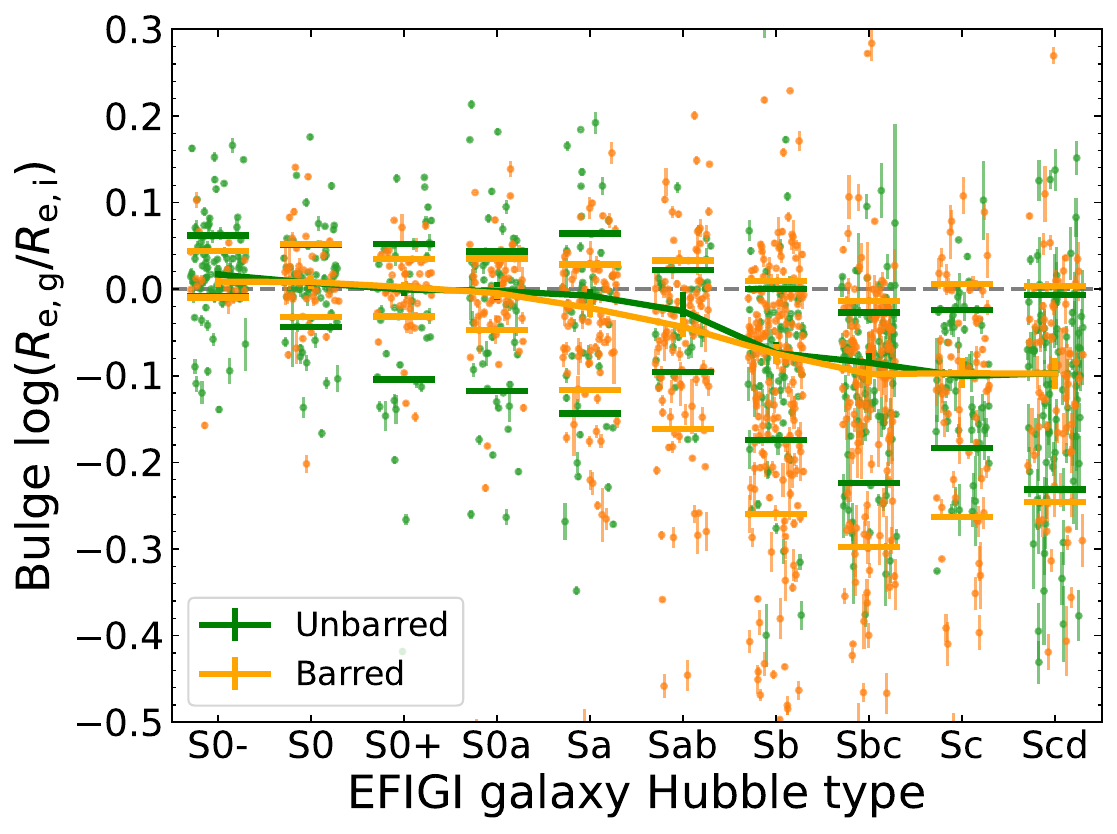}
\caption{Top: Distribution of the $g-i$ bulge colors for each Hubble type, similarly to the left panel of \fg\ref{fig-color-B-D}, separately for the barred (orange) and unbarred (green) samples. Bottom: Distribution of the $g$ to $i$ bulge $R_{\rm e}$ ratio (color gradient) per Hubble type, similarly to the left panel of \fg\ref{fig-color-gradients-bulges}, separately for the barred (orange) and unbarred samples (green). The same trends as a function of type are observed for both the barred and unbarred samples.}
\label{fig-impact-bar}
\end{figure*}

We mention in \sct\ref{sct-methodo} that the priors set on the S\'ersic profile fitting to the bulges are specifically designed to avoid fitting an elongated bulge component onto the bar, if present, as this obviously would bias the bulge measurements. To verify that the results presented in our analysis are not impacted by such biases, we show in \fg\ref{fig-impact-bar} the distribution of the $g-i$ bulge colors (left panel) and bulge color gradients as the $g$ to $i$ $R_{\rm e}$ ratio (right panel), separately for the unbarred (green) and barred (orange) samples: these subsamples are defined using the EFIGI visual attribute \texttt{Bar}, with respective values of $0$ and $1-4$ respectively (the different values signify the length of the bar, not its strength). Both panels show that the results obtained for the bulges of EFIGI galaxies described in the main body of the article remain true for both the barred and unbarred samples analyzed separately. This rules out that profile modeling biases due to the bars could cause the bulge color and gradient effects described in \sct\ref{sct-bulge-gradient}.

\section{Mass versus morphology\label{sct-appendix-mass}}

\begin{figure*}[ht]
\centering
\includegraphics[width=\columnwidth]{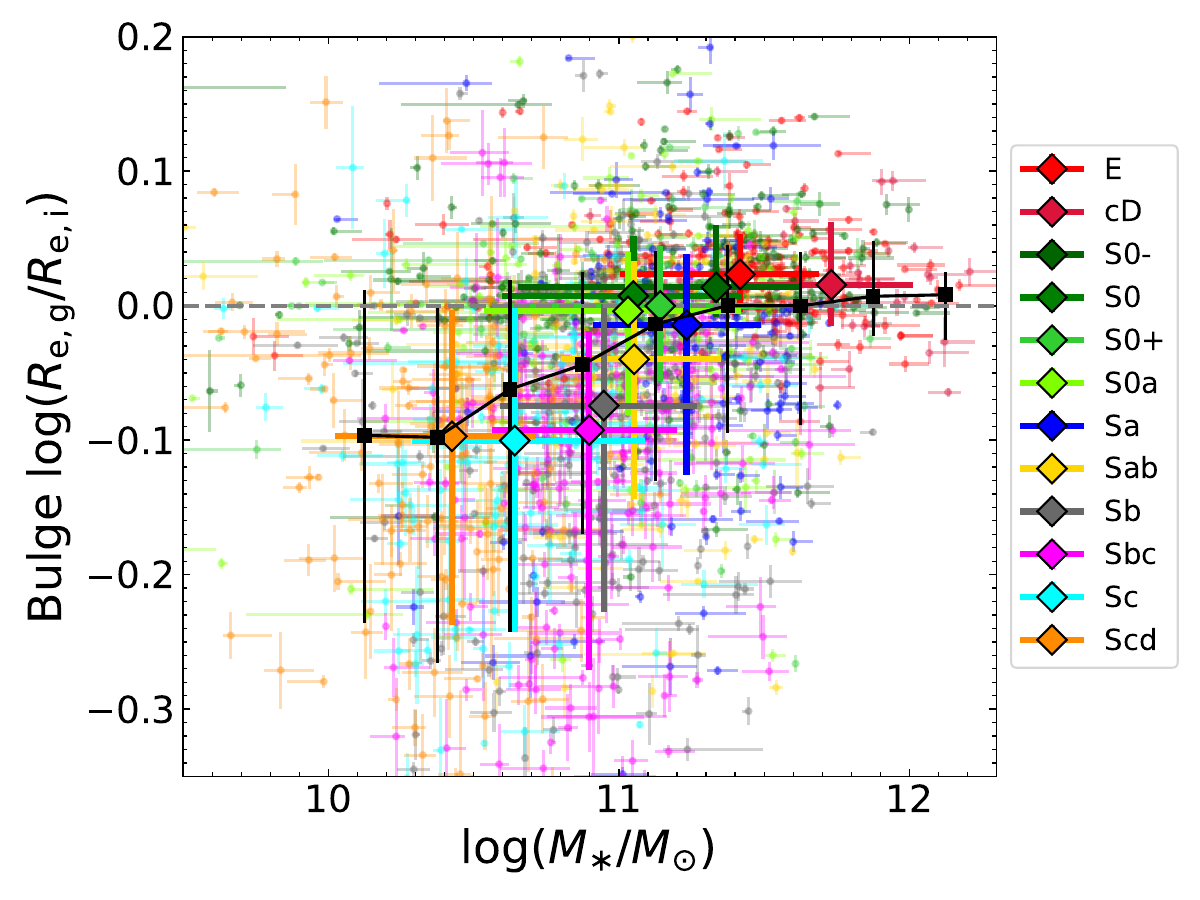}
\includegraphics[width=\columnwidth]{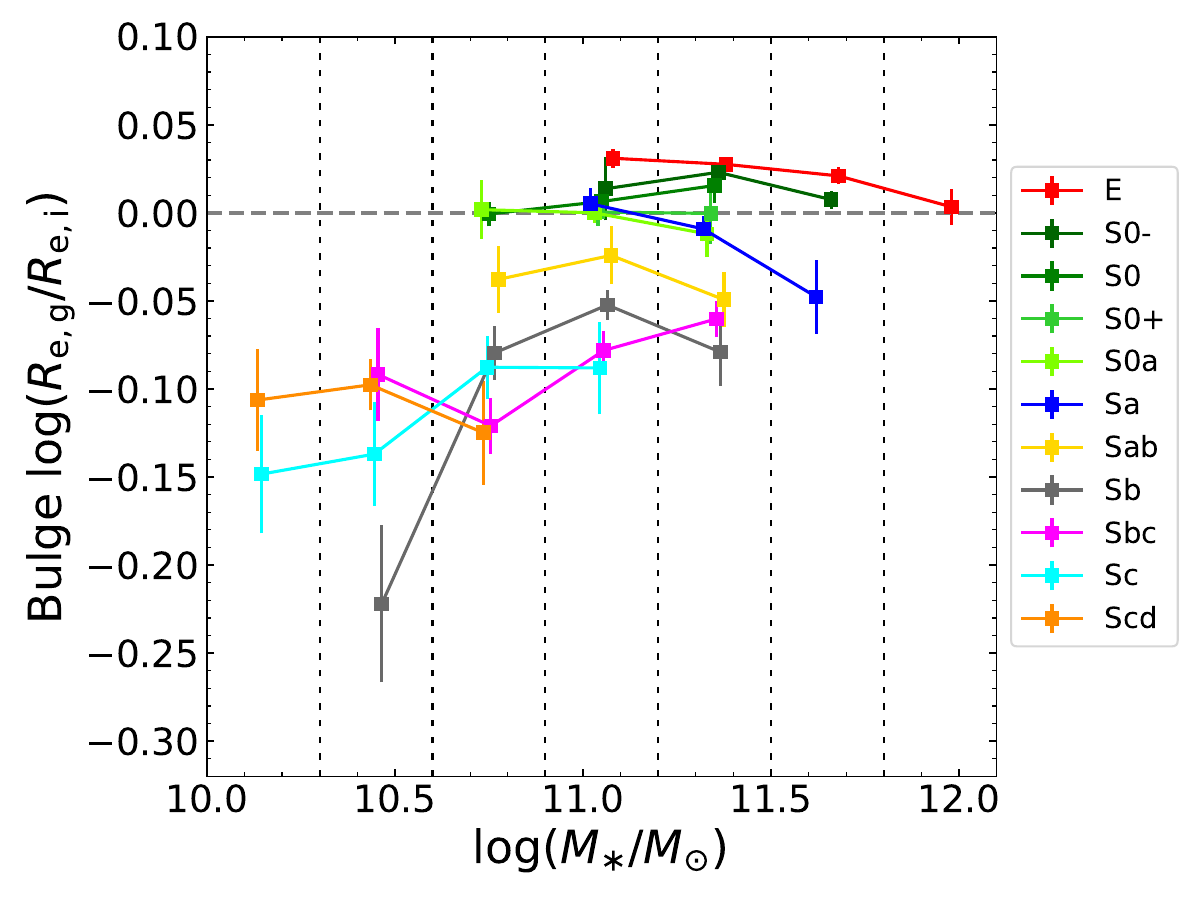}
\caption{Left: Ratios of bulge effective radii in $g$ over $i$ versus the total galaxy stellar masses, color-coded by their Hubble type. The median values for each Hubble type is indicated by a colored diamond, and the 16th to 84th percentiles by the horizontal and vertical bars. Black squares and their associated error bars indicate the medians and 16th to 84th percentiles, respectively, of the $R_e$ ratio per mass bin. Right: Median values (and shown with uncertainties in the median as determined using a bootstrap technique) of the bulge effective radii in $g$ over $i$ color gradient as a function of type, in intervals of total galaxy stellar mass of width 0.3 dex and requiring a minimum of 15 galaxies per bin. For clarity, we slightly offset the bin centers for the different galaxy types. The data are consistent with each Hubble type having a constant gradient as a function of stellar mass.
}
\label{fig-bulge-gradients-vs-stellar-mass} 
\end{figure*}

The left panel of \fg\ref{fig-bulge-gradients-vs-stellar-mass} shows that at galaxy stellar masses of $\sim 10^{10.6-11.5} M_\odot$, almost all morphological types, from intermediate-type spirals to lenticulars are represented. However, over this range of total stellar masses, we showed in \sct\ref{sct-bulge-gradient} that the EFIGI sample of galaxies shows very different bulge gradients depending on their type (gradients and mass medians per type are plotted in color).
Moreover, the left panel of \fg\ref{fig-bulge-gradients-vs-stellar-mass} also shows a progressive decrease in both the median and dispersion in the distribution of the bulge color gradients as a function of mass (plotted in black).
This illustrates how simply plotting the color gradient distributions of galaxy samples without detailed morphological information would give the false impression of a smooth fading of the gradients with increasing stellar mass. We remind that the EFIGI sample is a representative sample of morphological types, not stellar mass. In principle, it is therefore sensitive to the impact of morphology on gradients and colors, and perhaps less sensitive to variables impacted by stellar mass.

To further illustrate the variations of the bulge color gradients with mass versus morphology, in the right panel of \fg\ref{fig-bulge-gradients-vs-stellar-mass}, we split the median gradients per type into 7 stellar mass intervals with $0.3$ dex widths. One can see that there is no statistically significant mass dependence in the color gradient for each type. The spiral types which lie predominately in the Green Valley, the Sab and Sa Hubble types, have gradients between those of the Hubble types that lie in the Blue Cloud (Sb to Scd) and those that lie along the Red Sequence (E to S0+). None of the lenticular and E galaxy types have a statistically significant trend in their color gradients with total stellar mass. This provides additional support for the hypothesis that the Green Valley represents a meaningful transition region between the late- and early-type spiral galaxies, whereby the late-type spirals with relatively large gradients transition to lenticular galaxies with no color gradients and no mass dependence on their gradients. This finding indicates that the morphology of entire galaxies, hence the various characteristics of both their bulges and disks, play an important role in how galaxies evolve along the path of growing their bulges and quenching their disks. This evolution does occur as total stellar mass increases, but the present results show that to have a change in bulge properties (here, gradient) with mass, there must be a change in type.

We only show the behavior of the bulge gradients here, but similar arguments can be made regarding disk gradients, as the shift in their median values per type always occur for types Sb to S0, with a well populated common mass interval.

\section{Tables for $g-i$ and $g-r$ colors \label{sct-appendix-colors}}

We provide in Tables~\ref{tab-colors-gi} and \ref{tab-colors-gr} extensive statistics of the bulge, disk, and galaxy color distributions per morphological type, in $g-i$ and $g-r$, respectively. 

\begin{table*}[ht]
\caption{Statistics of the $g-i$ colors of the bulges, disks, whole galaxies, for all EFIGI morphological types.}
\resizebox{\linewidth}{!}{%
\begin{tabular}{  l r r r r r r r r r r r r r r r r }
\hline
\hline
\multicolumn{12}{c}{Bulge}\\
\hline
Type & cD & E & S0$^-$ & S0 & S0$^+$ & S0a & Sa & Sab & Sb & Sbc & Sc & Scd  \\
\hline
$N^{(a)}$ & 39 & 203 & 131 & 112 & 105 & 125 & 143 & 144 & 278 & 249 & 124 & 167 \\
Median & 1.021 & 1.007 & 0.996 & 1.003 & 1.022 & 1.044 & 1.104 & 1.124 & 1.182 & 1.238 & 1.153 & 1.057 \\
Error$^{(b)}$ & 0.019 & 0.005 & 0.006 & 0.012 & 0.02 & 0.022 & 0.018 & 0.012 & 0.020 & 0.019 & 0.018 & 0.033 \\
${q_{16}}^{(c)}$ & 0.957 & 0.924 & 0.912 & 0.904 & 0.846 & 0.907 & 0.973 & 0.933 & 0.955 & 1.054 & 0.879 & 0.714 \\
${q_{84}}^{(c)}$ & 1.076 & 1.074 & 1.063 & 1.082 & 1.227 & 1.219 & 1.261 & 1.304 & 1.464 & 1.511 & 1.402 & 1.369 \\
Mean & 1.019 & 1.000 & 0.987 & 0.995 & 1.021 & 1.047 & 1.106 & 1.117 & 1.192 & 1.255 & 1.142 & 1.051 \\
$\mathrm{SD}^{(d)}$ & 0.092 & 0.082 & 0.113 & 0.175 & 0.286 & 0.233 & 0.204 & 0.230 & 0.303 & 0.328 & 0.296 & 0.342 \\
\hline
\multicolumn{16}{c}{Disk}\\
\hline
Type & cD & E & S0$^-$ & S0 & S0$^+$ & S0a & Sa & Sab & Sb & Sbc & Sc & Scd & Sd$^{(e)}$ & Sdm$^{(e)}$ & Sm$^{(e)}$ & Im$^{(e)}$ \\
\hline
$N^{(a)}$ & 39 & 203 & 131 & 112 & 105 & 125 & 143 & 144 & 278 & 249 & 124 & 167 & 122 & 211 & 183 & 132 \\
Median & 1.012 & 1.043 & 1.009 & 0.972 & 0.948 & 0.876 & 0.863 & 0.75 & 0.683 & 0.644 & 0.624 & 0.502 & 0.421 & 0.400 & 0.405 & 0.337 \\
Error$^{(b)}$ & 0.013 & 0.007 & 0.005 & 0.014 & 0.018 & 0.016 & 0.011 & 0.013 & 0.012 & 0.007 & 0.019 & 0.013 & 0.017 & 0.013 & 0.016 & 0.016 \\
${q_{16}}^{(c)}$ & 0.951 & 0.960 & 0.903 & 0.893 & 0.825 & 0.673 & 0.738 & 0.615 & 0.545 & 0.525 & 0.464 & 0.378 & 0.305 & 0.276 & 0.258 & 0.189 \\
${q_{84}}^{(c)}$ & 1.090 & 1.111 & 1.081 & 1.059 & 1.041 & 0.982 & 0.963 & 0.845 & 0.819 & 0.745 & 0.724 & 0.615 & 0.519 & 0.532 & 0.608 & 0.499 \\
Mean & 1.008 & 1.036 & 0.992 & 0.974 & 0.939 & 0.849 & 0.857 & 0.734 & 0.68 & 0.635 & 0.607 & 0.502 & 0.423 & 0.408 & 0.44 & 0.349 \\
$\mathrm{SD}^{(d)}$ & 0.091 & 0.086 & 0.101 & 0.109 & 0.130 & 0.169 & 0.140 & 0.111 & 0.131 & 0.121 & 0.137 & 0.128 & 0.118 & 0.149 & 0.223 & 0.193 \\
\hline
\multicolumn{16}{c}{Galaxy$^{(e)}$}\\
\hline
Type & cD & E & S0$^-$ & S0 & S0$^+$ & S0a & Sa & Sab & Sb & Sbc & Sc & Scd & Sd$^{(e)}$ & Sdm$^{(e)}$ & Sm$^{(e)}$ & Im$^{(e)}$ \\
\hline
$N^{(a)}$ & 39 & 203 & 131 & 112 & 105 & 125 & 143 & 144 & 278 & 249 & 124 & 167 & 122 & 211 & 183 & 132 \\
Median & 0.999 & 1.015 & 0.996 & 0.978 & 0.995 & 0.930 & 0.931 & 0.842 & 0.759 & 0.701 & 0.658 & 0.521 & 0.421 & 0.400 & 0.405 & 0.337 \\
Error$^{(b)}$ & 0.006 & 0.004 & 0.008 & 0.008 & 0.009 & 0.015 & 0.007 & 0.011 & 0.011 & 0.008 & 0.014 & 0.011 & 0.017 & 0.013 & 0.016 & 0.015 \\
${q_{16}}^{(c)}$ & 0.966 & 0.964 & 0.917 & 0.924 & 0.857 & 0.824 & 0.835 & 0.728 & 0.627 & 0.565 & 0.498 & 0.390 & 0.305 & 0.276 & 0.258 & 0.189 \\
${q_{84}}^{(c)}$ & 1.047 & 1.060 & 1.036 & 1.033 & 1.045 & 1.011 & 1.001 & 0.921 & 0.889 & 0.803 & 0.745 & 0.641 & 0.519 & 0.532 & 0.608 & 0.499 \\
Mean & 0.998 & 1.009 & 0.979 & 0.970 & 0.968 & 0.914 & 0.915 & 0.825 & 0.756 & 0.689 & 0.643 & 0.524 & 0.423 & 0.408 & 0.44 & 0.349 \\
$\mathrm{SD}^{(d)}$ & 0.044 & 0.054 & 0.075 & 0.086 & 0.117 & 0.120 & 0.094 & 0.097 & 0.124 & 0.115 & 0.134 & 0.127 & 0.118 & 0.149 & 0.223 & 0.193 \\
\hline
\multicolumn{12}{c}{Bulge$-$Disk$^{(f)}$}\\
\hline
Type & cD & E & S0$^-$ & S0 & S0$^+$ & S0a & Sa & Sab & Sb & Sbc & Sc & Scd \\
\hline
$N^{(a)}$ & 39 & 203 & 131 & 112 & 105 & 125 & 143 & 144 & 278 & 249 & 124 & 167 \\
Median & 0.011 & -0.036 & -0.005 & 0.016 & 0.074 & 0.205 & 0.231 & 0.378 & 0.489 & 0.606 & 0.536 & 0.552 \\
Error$^{(b)}$ & 0.019 & 0.009 & 0.010 & 0.010 & 0.026 & 0.019 & 0.018 & 0.013 & 0.022 & 0.027 & 0.027 & 0.027 \\
${q_{16}}^{(c)}$ & -0.129 & -0.150 & -0.119 & -0.107 & -0.100 & 0.005 & 0.024 & 0.173 & 0.240 & 0.354 & 0.253 & 0.294 \\
${q_{84}}^{(c)}$ & 0.109 & 0.078 & 0.091 & 0.142 & 0.360 & 0.441 & 0.436 & 0.618 & 0.810 & 0.908 & 0.858 & 0.824 \\
Mean & 0.011 & -0.036 & -0.005 & 0.021 & 0.081 & 0.198 & 0.249 & 0.383 & 0.511 & 0.620 & 0.535 & 0.549 \\
$\mathrm{SD}^{(d)}$ & 0.154 & 0.123 & 0.148 & 0.198 & 0.308 & 0.282 & 0.271 & 0.258 & 0.322 & 0.354 & 0.301 & 0.318 \\
\hline
\end{tabular}
}
\small\textbf{Notes.}
$^{(a)}$ Number of galaxies per Hubble type in which both the bulge and disk S\'ersic fits, for galaxies with Scd types and earlier, or the single S\'ersic fits for Sd through Im types, are assessed as reliable. 
$^{(b)}$ Error in the median $g-i$ color per Hubble type, determined by bootstrapping.
$^{(c)}$ 16\% and 84\% percentiles of the $g-i$ distribution per Hubble type.
$^{(d)}$ Standard deviation of $g-i$ per Hubble type.
$^{(e)}$ The single-S\'ersic models of Sd-type and later are listed as disks because they do not host a significant bulge and their S\'ersic index distribution peaks around 1. The statistics of the single-profile fits are enumerated in both the disk and whole galaxy.
\label{tab-colors-gi}
\end{table*}

\begin{table*}[ht]
\caption{Statistics of the $g-r$ colors of the bulges, disks, whole galaxies, for all EFIGI morphological types.}
\resizebox{\linewidth}{!}{%
\begin{tabular}{  l r r r r r r r r r r r r r r r r }
\hline
\hline
\multicolumn{12}{c}{Bulge}\\
\hline
Type & cD & E & S0$^-$ & S0 & S0$^+$ & S0a & Sa & Sab & Sb & Sbc & Sc & Scd  \\
\hline
$N^{(a)}$ & 39 & 203 & 131 & 112 & 105 & 125 & 143 & 144 & 278 & 249 & 124 & 167 \\
Median & 0.696 & 0.680 & 0.672 & 0.671 & 0.689 & 0.709 & 0.747 & 0.747 & 0.781 & 0.815 & 0.743 & 0.684 \\
Error$^{(b)}$ & 0.013 & 0.005 & 0.006 & 0.007 & 0.013 & 0.008 & 0.009 & 0.009 & 0.013 & 0.012 & 0.013 & 0.020 \\
${q_{16}}^{(c)}$ & 0.605 & 0.622 & 0.619 & 0.604 & 0.561 & 0.600 & 0.638 & 0.625 & 0.629 & 0.684 & 0.567 & 0.505 \\
${q_{84}}^{(c)}$ & 0.738 & 0.732 & 0.729 & 0.757 & 0.826 & 0.797 & 0.858 & 0.902 & 0.999 & 0.990 & 0.938 & 0.873 \\
Mean & 0.679 & 0.681 & 0.671 & 0.671 & 0.674 & 0.697 & 0.757 & 0.750 & 0.794 & 0.831 & 0.746 & 0.678 \\
$\mathrm{SD}^{(d)}$ & 0.070 & 0.067 & 0.081 & 0.122 & 0.199 & 0.153 & 0.171 & 0.154 & 0.234 & 0.230 & 0.241 & 0.271 \\
\hline
\multicolumn{16}{c}{Disk}\\
\hline
Type & cD & E & S0$^-$ & S0 & S0$^+$ & S0a & Sa & Sab & Sb & Sbc & Sc & Scd & Sd$^{(e)}$ & Sdm$^{(e)}$ & Sm$^{(e)}$ & Im$^{(e)}$ \\
\hline
$N^{(a)}$ & 39 & 203 & 131 & 112 & 105 & 125 & 143 & 144 & 278 & 249 & 124 & 167 & 122 & 211 & 183 & 132 \\
Median & 0.633 & 0.631 & 0.627 & 0.602 & 0.598 & 0.559 & 0.542 & 0.478 & 0.441 & 0.417 & 0.398 & 0.33 & 0.265 & 0.25 & 0.251 & 0.223 \\
Error$^{(b)}$ & 0.01 & 0.005 & 0.007 & 0.007 & 0.013 & 0.01 & 0.007 & 0.008 & 0.008 & 0.006 & 0.012 & 0.008 & 0.01 & 0.007 & 0.008 & 0.01 \\
${q_{16}}^{(c)}$ & 0.552 & 0.555 & 0.563 & 0.552 & 0.509 & 0.436 & 0.456 & 0.37 & 0.347 & 0.336 & 0.287 & 0.244 & 0.191 & 0.183 & 0.175 & 0.115 \\
${q_{84}}^{(c)}$ & 0.686 & 0.686 & 0.682 & 0.672 & 0.665 & 0.617 & 0.608 & 0.538 & 0.528 & 0.486 & 0.468 & 0.4 & 0.352 & 0.346 & 0.384 & 0.322 \\
Mean & 0.627 & 0.624 & 0.62 & 0.605 & 0.596 & 0.54 & 0.537 & 0.46 & 0.435 & 0.408 & 0.389 & 0.325 & 0.266 & 0.26 & 0.269 & 0.232 \\
$\mathrm{SD}^{(d)}$ & 0.071 & 0.078 & 0.072 & 0.07 & 0.098 & 0.132 & 0.102 & 0.078 & 0.086 & 0.079 & 0.087 & 0.084 & 0.092 & 0.125 & 0.125 & 0.128 \\
\hline
\multicolumn{16}{c}{Galaxy$^{(e)}$}\\
\hline
Type & cD & E & S0$^-$ & S0 & S0$^+$ & S0a & Sa & Sab & Sb & Sbc & Sc & Scd & Sd$^{(e)}$ & Sdm$^{(e)}$ & Sm$^{(e)}$ & Im$^{(e)}$ \\
\hline
$N^{(a)}$ & 39 & 203 & 131 & 112 & 105 & 125 & 143 & 144 & 278 & 249 & 124 & 167 & 122 & 211 & 183 & 132 \\
Median & 0.644 & 0.65 & 0.645 & 0.635 & 0.642 & 0.604 & 0.601 & 0.532 & 0.495 & 0.449 & 0.419 & 0.34 & 0.265 & 0.25 & 0.251 & 0.223 \\
Error$^{(b)}$ & 0.008 & 0.003 & 0.005 & 0.007 & 0.006 & 0.008 & 0.005 & 0.008 & 0.007 & 0.006 & 0.012 & 0.008 & 0.011 & 0.007 & 0.008 & 0.009 \\
${q_{16}}^{(c)}$ & 0.607 & 0.611 & 0.592 & 0.59 & 0.551 & 0.522 & 0.532 & 0.464 & 0.412 & 0.374 & 0.323 & 0.253 & 0.191 & 0.183 & 0.175 & 0.115 \\
${q_{84}}^{(c)}$ & 0.663 & 0.685 & 0.676 & 0.678 & 0.684 & 0.661 & 0.644 & 0.588 & 0.562 & 0.512 & 0.478 & 0.416 & 0.352 & 0.346 & 0.384 & 0.322 \\
Mean & 0.638 & 0.648 & 0.636 & 0.629 & 0.625 & 0.592 & 0.589 & 0.527 & 0.487 & 0.442 & 0.411 & 0.338 & 0.266 & 0.26 & 0.269 & 0.232 \\
$\mathrm{SD}^{(d)}$ & 0.032 & 0.036 & 0.048 & 0.056 & 0.074 & 0.074 & 0.066 & 0.063 & 0.077 & 0.073 & 0.081 & 0.083 & 0.092 & 0.125 & 0.125 & 0.128 \\
\hline
\multicolumn{12}{c}{Bulge$-$Disk$^{(f)}$}\\
\hline
Type & cD & E & S0$^-$ & S0 & S0$^+$ & S0a & Sa & Sab & Sb & Sbc & Sc & Scd \\
\hline
$N^{(a)}$ & 39 & 203 & 131 & 112 & 105 & 125 & 143 & 144 & 278 & 249 & 124 & 167 \\
Median & 0.059 & 0.048 & 0.041 & 0.058 & 0.076 & 0.173 & 0.212 & 0.271 & 0.344 & 0.403 & 0.351 & 0.356 \\
Error$^{(b)}$ & 0.013 & 0.007 & 0.01 & 0.008 & 0.022 & 0.016 & 0.014 & 0.013 & 0.012 & 0.016 & 0.02 & 0.018 \\
${q_{16}}^{(c)}$ & -0.038 & -0.036 & -0.02 & -0.025 & -0.034 & 0.024 & 0.043 & 0.138 & 0.181 & 0.248 & 0.164 & 0.198 \\
${q_{84}}^{(c)}$ & 0.133 & 0.15 & 0.121 & 0.188 & 0.256 & 0.323 & 0.386 & 0.455 & 0.575 & 0.61 & 0.55 & 0.554 \\
Mean & 0.052 & 0.057 & 0.051 & 0.065 & 0.078 & 0.157 & 0.219 & 0.29 & 0.36 & 0.423 & 0.357 & 0.353 \\
$\mathrm{SD}^{(d)}$ & 0.119 & 0.116 & 0.113 & 0.142 & 0.238 & 0.21 & 0.223 & 0.182 & 0.252 & 0.245 & 0.258 & 0.251 \\
\hline
\end{tabular}
}
\small\textbf{Notes.}
$^{(a)}$ Number of galaxies per Hubble type in which both the bulge and the disk fits (for Scd types and earlier), or the single S\'ersic fits (for Sd to Im types) are assessed as reliable. 
$^{(b)}$ Error in the median $g-r$ color per Hubble type, determined by bootstrapping.
$^{(c)}$ 16\% and 84\% percentiles of the $g-r$ distribution per Hubble type.
$^{(d)}$ Standard deviation of $g-r$ per Hubble type.
$^{(e)}$ The single-S\'ersic models of Sd-type and later are listed as disks because they do not host a significant bulge and their S\'ersic index distribution peaks around 1. The statistics of the single-profile fits are listed for both the disk and whole galaxy.
\label{tab-colors-gr}
\end{table*}

\section{Tables for the $R_{\rm e}$ variations with bands\label{sct-appendix-gradients}}

In this section we provide extensive statistics for the bulge and disk color gradients, as measured by the log-ratio between their $g$ and $i$-band $R_{\rm e}$ distributions per morphological type.

\begin{table*}[ht]
\caption{Statistics of the ratios of effective radii in the $g$ and $i$ bands, for the bulges and disks of the EFIGI morphological types.}
\resizebox{\linewidth}{!}{%
\begin{tabular}{  l r r r r r r r r r r r r r r r r }
\hline
\hline
\multicolumn{16}{c}{Disk}\\
\hline
Type & cD & E & S0$^-$ & S0 & S0$^+$ & S0a & Sa & Sab & Sb & Sbc & Sc & Scd & Sd$^{(a)}$ & Sdm$^{(a)}$ & Sm$^{(a)}$ & Im$^{(a)}$ \\
\hline
$N^{(b)}$ & 39 & 203 & 131 & 112 & 105 & 125 & 143 & 144 & 278 & 249 & 124 & 167 & 122 & 211 & 183 & 132 \\
Median & 0.017 & 0.010 & 0.010 & 0.000 & 0.002 & 0.007 & 0.018 & 0.033 & 0.027 & 0.045 & 0.042 & 0.046 & -0.003 & 0.012 & 0.001 & -0.006 \\
Error & 0.005 & 0.002 & 0.003 & 0.002 & 0.002 & 0.004 & 0.003 & 0.003 & 0.002 & 0.002 & 0.003 & 0.002 & 0.005 & 0.005 & 0.004 & 0.006 \\
$q_{16}$ & -0.013 & -0.011 & -0.013 & -0.017 & -0.025 & -0.024 & -0.011 & -0.007 & 0.001 & 0.022 & 0.019 & 0.018 & -0.053 & -0.041 & -0.032 & -0.044 \\
$q_{84}$ & 0.055 & 0.037 & 0.040 & 0.025 & 0.022 & 0.047 & 0.049 & 0.064 & 0.057 & 0.07 & 0.067 & 0.063 & 0.072 & 0.098 & 0.047 & 0.034 \\
Mean & 0.018 & 0.014 & 0.013 & 0.005 & -0.001 & 0.015 & 0.02 & 0.031 & 0.027 & 0.046 & 0.043 & 0.043 & 0.007 & 0.022 & 0.003 & -0.002 \\
$\mathrm{SD}$ & 0.052 & 0.030 & 0.034 & 0.039 & 0.033 & 0.045 & 0.038 & 0.039 & 0.031 & 0.027 & 0.025 & 0.026 & 0.074 & 0.078 & 0.068 & 0.084 \\
Median ${R_{{\rm e}, g}}^{(c)}$ & 24.6 & 11.2 & 9.9 & 6.4 & 6.7 & 7.9 & 9.6 & 9.5 & 7.7 & 9.5 & 8.7 & 8.1 & 6.7 & 6.4 & 3.2 & 2.1 \\
\hline
\multicolumn{12}{c}{Bulge}\\
\hline
Type & cD & E & S0$^-$ & S0 & S0$^+$ & S0a & Sa & Sab & Sb & Sbc & Sc & Scd \\
\hline
$N^{(b)}$ & 39 & 203 & 131 & 112 & 105 & 125 & 143 & 144 & 278 & 249 & 124 & 167 &   &   &   &   \\
Median & 0.016 & 0.023 & 0.014 & 0.007 & 0.000 & -0.004 & -0.014 & -0.040 & -0.074 & -0.093 & -0.100 & -0.097 &   &   &   &   \\
Error & 0.007 & 0.003 & 0.004 & 0.005 & 0.002 & 0.005 & 0.007 & 0.010 & 0.008 & 0.008 & 0.010 & 0.011 &   &   &   &   \\
$q_{16}$ & -0.015 & -0.005 & -0.009 & -0.041 & -0.056 & -0.081 & -0.126 & -0.144 & -0.226 & -0.271 & -0.242 & -0.238 &   &   &   &   \\
$q_{84}$ & 0.062 & 0.053 & 0.060 & 0.052 & 0.044 & 0.04 & 0.039 & 0.033 & 0.001 & -0.019 & -0.002 & -0.003 &   &   &   &   \\
Mean & 0.016 & 0.026 & 0.019 & 0.005 & -0.009 & -0.022 & -0.038 & -0.057 & -0.105 & -0.122 & -0.112 & -0.112 &   &   &   &   \\
$\mathrm{SD}$ & 0.045 & 0.038 & 0.052 & 0.057 & 0.082 & 0.101 & 0.103 & 0.112 & 0.148 & 0.168 & 0.142 & 0.131 &   &   &   &   \\
Median ${R_{{\rm e}, g}}^{(c)}$ & 5.0 & 2.3 & 1.9 & 1.0 & 1.3 & 1.2 & 1.2 & 0.9 & 0.7 & 0.7 & 0.5 & 0.6 &   &   &   &   \\
\hline
\end{tabular}
}
\small\textbf{Notes.}
$^{(a)}$ We list the statistics of the single-S\'ersic profile modeling of Sd-type and later as disks because they do not host a significant bulge and their S\'ersic index distributions peak around 1. 
$^{(b)}$ Number of galaxies per Hubble type in which both the bulge and the disk fits (for Scd types and earlier), or the single S\'ersic fits (for Sd to Im types) are assessed as reliable.
$^{(c)}$ The median bulge or disk effective radii in kpc. 
\label{tab-gradients-gi}
\end{table*}

\end{appendix}

\end{document}